\documentclass[%
 reprint,
 amsmath,amssymb,
 aps,
]{revtex4-1}

\usepackage{graphicx}
\usepackage{dcolumn}
\usepackage{bm}
\graphicspath{{fig/}}
\usepackage[colorlinks=true, linkcolor=black, citecolor=blue, urlcolor=blue]{hyperref} 

\usepackage{enumitem}
\setlist{  
  listparindent=\parindent,
  parsep=0pt,
}

\newcommand{\trento}{T\raisebox{-.5ex}{R}ENTo }

\begin{document}

\title{A data-driven analysis for the temperature and momentum dependence of the heavy quark diffusion coefficient in relativistic heavy-ion collisions}

\author{Yingru Xu}
\email[Correspond to\ ]{yx59@phy.duke.edu}
\affiliation{Department of Physics, Duke University, Durham, NC 27708, USA}

\author{Marlene Nahrgang}
\affiliation{SUBATECH, UMR 6457, IMT Atlantique, Universit´e de Nantes, IN2P3/CNRS, 4 rue Alfred Kastler, 44307 Nantes cedex 3, France}

\author{Shanshan Cao}
\affiliation{Department of Physics and Astronomy, Wayne State University, Detroit, MI, 48201}

\author{Jonah E. Bernhard}
\affiliation{Department of Physics, Duke University, Durham, NC 27708, USA}

\author{Steffen A. Bass}
\affiliation{Department of Physics, Duke University, Durham, NC 27708, USA}

\date{\today}

\begin{abstract}
By applying a Bayesian model-to-data analysis, we estimate the temperature and momentum dependence of the heavy quark diffusion coefficient in an improved Langevin framework. The posterior range of the diffusion coefficient is obtained by performing a Markov chain Monte Carlo random walk and calibrating on the experimental data of $D$-meson $R_{\mathrm{AA}}$ and $v_2$ in three different collision systems at RHIC and the LHC: AuAu collisions at 200 GeV, PbPb collisions at 2.76 and 5.02 TeV. The spatial diffusion coefficient is found to be consistent with lattice QCD calculations and comparable with other models' estimation. We demonstrate the capability of our improved Langevin model to simultaneously describe the $R_{\mathrm{AA}}$ and $v_2$ at both RHIC and the LHC energies, as well as the higher order flow coefficient such as $D$-meson $v_3$. We show that by applying a Bayesian analysis, we are able to quantitatively and systematically study the heavy flavor dynamics in heavy-ion collisions.
\end{abstract}

\maketitle

\section{\label{sec:intro}Introduction}
The theory of the strong interaction force -- Quantum Chromodynamics (QCD) -- predicts that at sufficiently high temperature and/or baryon density, nuclear matter undergoes a phase transition from hadrons to a new state of the deconfined quarks and gluons: the quark gluon plasma (QGP)~\cite{Bjorken:1982qr,Shuryak:1980tp, McLerran:1986zb}. Over the past two decades, ultra-relativistic heavy-ion collision experiments at the Relativistic Heavy Ion Collider (RHIC) and the Large Hadron Collider (LHC) have been searching and exploring this new state of matter under extreme conditions. Compelling discoveries, for instance the strong suppression of hadrons at large transverse momenta (jet quenching), reveal the creation of the QGP medium at RHIC and the LHC~\cite{Teaney:2000cw, Adler:2005ee}. The observed collective flow of low transverse momentum hadrons~\cite{Ackermann:2000tr,Abelev:2006db} has provided insight into remarkable properties of the QGP such as the strongly interacting, almost perfect fluid behavior with a very small shear viscosity to entropy density ratio~\cite{Song:2008hj,Son:2007vk,Kovtun:2004de}.

Since the QGP is not directly observable, the study of its properties relies on the measurement of final state observables, as well as theoretical modeling, and the comparison between those two. For example, the relativistic viscous hydrodynamical model~\cite{Luzum:2008cw,Heinz:2005bw,Schenke:2010rr,Bozek:2011ua,Karpenko:2013wva} -- one of the most successful models in heavy-ion physics -- has been utilized for the extraction of the temperature dependence of the specific shear-viscosity through a model-to-data comparison with elliptic and triangular flow data of soft identified hadrons~\cite{Bernhard:2016tnd}.

In contrast to the soft medium properties, the transport coefficients related to the medium interaction of hard probes (jets and heavy quarks), such as $\hat{q}$ and $\hat{e}, D_s$,  are not yet understood on a similarly quantitative level. This is in part due to the experimental difficulty in measuring  ``rare processes'', but also due to the complexity of modeling the dynamics of these hard probes interacting with the QGP medium. Nevertheless, significant progress has been made in recent years: a number of transport models in the market are now able to describe a selection of heavy quark observables and perform qualitative estimates of the diffusion coefficient~\cite{Wicks:2007am,Sharma:2009hn,Nahrgang:2013saa,Uphoff:2010sh,He:2011yi,Cao:2013ita,Song:2015ykw,CaronHuot:2008uh, He:2012df,vanHees:2007me, Scardina:2017ipo,Riek:2010fk,Gossiaux:2008jv,Cao:2016gvr}, which in turn can be compared with lattice QCD calculations of the same quantities~\cite{Ding:2012sp,Francis:2015daa,Banerjee:2011ra}.
In current studies of the open heavy flavor diffusion coefficient, it is common that the diffusion coefficient is directly or indirectly encoded in the model and one can relate its physical properties to one or multiple parameters. By comparing the heavy quark observables (such as the nuclear modification factor $R_{\mathrm{AA}}$ and elliptic flow $v_2$) between the theoretical calculation and the experimental data, these parameters can be tuned until one finds a satisfactory fit. However, the disadvantage of such an ``eyeball'' comparison is that it gets exceedingly difficult to vary multiple parameters simultaneously or to compare with a larger selection of experimental measurements, as all parameters are interdependent and affect multiple observables at once. 

A more rigorous and complete approach to optimizing the model and determining the parameters would be to perform a random walk in the parameter space and calibrate on the experimental data by applying a modern Bayesian statistical analysis~\cite{Habib:2007ca, Higdon:2014tva,Higdon:2008cmc,Novak:2013bqa}. In such an analysis, the computationally expensive physics model is first evaluated for a small number of points in parameter space. These calculations are used to train Gaussian Process emulators that act as fast surrogates to interpolate between these points and provide model predictions for arbitrary values of the input parameters. The emulators thus act as substitution of the full model in order to be able to perform a Markov chain Monte Carlo exploration of the complete parameter space. The results of such an analysis are the posterior distributions of the parameters, i.e. the probability distributions of the parameter values that optimally describe the experimental data.

This type of model-to-data analysis using  Bayesian statistics has been applied with great success in the soft sector of heavy-ion physics over the last few years, e.g. for the extraction of the temperature dependence of the specific shear and bulk viscosities of the QGP~\cite{Bernhard:2016tnd,Auvinen:2017fjw},  as well as for constraining  the equation of state of QCD matter purely from experimental measurements~\cite{Novak:2013bqa,Pratt:2015zsa}. In this study, we shall extend this type of analysis to the quantitative study of heavy flavor evolution in heavy-ion collisions. Our goal is to provide a quantitative estimate of the temperature and momentum dependence of the heavy flavor diffusion coefficient. The paper is organized as follows: in Sec.~\ref{sec:model} we will describe our physics model, the improved Langevin framework which simulates the full space-time evolution of heavy quarks inside a QGP medium; in Sec.~\ref{sec:Bayesian} the Bayesian analysis that we utilize is introduced; the results of the calibration and the estimation of the heavy quark diffusion coefficient are given in Sec.~\ref{sec:results}; a summary and outlook can be found in the final section.

\section{\label{sec:model} Modeling heavy flavor evolution in heavy-ion collisions}
\subsection{\label{subsec:overall-model}Full space-time evolution of heavy flavors and the QGP medium}

Our analysis utilizes the well-established framework developed by the Duke QCD group to simulate the full space-time evolution of heavy quarks in heavy-ion collisions~\cite{Cao:2013ita,Cao:2015hia}:

\subsubsection{Initial production}

Due to their large masses, heavy quarks are believed to be primarily produced at the beginning of the collision via hard scattering. Therefore the initial momentum distribution can be calculated using perturbative QCD (pQCD). In this work, we adopt the fixed-order plus next-to-leading log formula (FONLL)~\cite{Cacciari:1998it,Cacciari:2001td} and EPS09 NLO nuclear PDFs~\cite{Eskola:2009uj} to calculate the heavy quark initial momentum distribution, from which we sample the momenta of heavy quarks in our calculation (using Monte Carlo methods).

The initial distribution of heavy quarks in position space is generated consistently with the initial condition for the event-by-event hydrodynamical evolution by the parametric initial condition model \trento\cite{Moreland:2014oya}. At the soft medium thermalization time ($\tau_0=0.6$ fm/c), \trento maps the entropy density $s(x,y)|_{\tau_0}$ to the nucleon thickness functions $T_A, T_B$ of the two projectiles by asserting a generalized ansatz:
\begin{eqnarray}\label{eqn:trento}
s(x,y)|_{\tau=\tau_0} \propto \left(\frac{T_A^p + T_B^p}{2}\right)^{1/p}.
\end{eqnarray}
where $p$ is a free parameter in \trento. In this study, we utilize the calibration of the soft matter properties performed in~\cite{Bernhard:2016tnd} (Table III of the median value calibrated on charged particle yields of~\cite{Bernhard:2016tnd}), including \trento initial state parameters (except for $p$, which is chosen to be 0), as well as other parameters related to soft medium properties, therefore the generalized function can be simplified as
\begin{equation}
s(x,y)|_{\tau=\tau_0} = \sqrt{T_AT_B}.
\end{equation}
The heavy quark initial production is based on the binary collision scaling and is determined by the thickness function $\hat{T}_{AB}=T_AT_B$. In this way, we are able to relate the heavy quark initial position to the soft medium initial production.

\subsubsection{\label{sec:Langevin}Heavy quark evolution inside a QGP medium} 

After their production, heavy quarks propagate through the QGP medium. In the quasi-particle picture of the QGP system, the space-time evolution of heavy quarks can generally be described by the Boltzmann equation. Since heavy quark masses are much larger than the typical medium temperature ($m_Q\gg T$), their momentum change due to the scattering with light partons in a thermally equilibrated medium is relatively small. With this assumption, the Boltzmann equation can be reduced to the Fokker-Planck equation, which is realized stochastically by the Langevin equation (for a detailed derivation, see Refs.~\cite{Rapp:2009my,Moore:2004tg}). In this study, we use an improved Langevin transport model to describe the dynamics of heavy quarks propagating in a QGP medium, which includes not only the heavy quark collisional energy loss but also the radiative energy loss due to gluon radiation~\cite{Cao:2013ita}
\begin{equation}
\frac{d\vec{p}}{dt} = -\eta_D(p)\vec{p} + \vec{\xi} + \vec{f}_g.
\end{equation}
The first two terms on the right hand side of the equation are the drag and thermal random forces inherited from the standard Langevin equation. They contribute to the collisional energy loss from quasi-elastic scattering between heavy quarks and light partons, and are generalized to the scattering between heavy quarks and the background medium. With the requirement that the heavy quark distribution eventually reaches equilibrium in a thermal medium, a simplified form of the Einstein relation $\eta_D(p) = \kappa /(2 TE)$ is adopted, where $\kappa$ denotes the heavy quark mean squared momentum change per unit time, and is usually referred as momentum-space diffusion coefficient. Here we assume that in a ``minimal model'', the heavy quark momentum variance in the longitudinal direction equals that in the transverse direction (even though the microscopic calculation of those two quantities are different): $\kappa_{||} = \kappa_{\perp} = \kappa$. The validity of such an assumption needs to be investigated in a future calculation. For this study we follow this assumption in order to simplify our parameterization, and therefore the heavy quark transport coefficient is defined as $\hat{q} = 2\kappa_{\perp}$, which is the transverse momentum broadening.

Assuming Gaussian shaped white noise, the thermal random force satisfies the relation $\left<\xi_i(t) \xi_j(t')\right> = \kappa \delta_{ij}\delta(t-t')$, which indicates no correlation between thermal forces at different times.

In order to describe the heavy quark dynamics in the intermediate and high $p_{\mathrm{T}}$ region, the effective modeling the radiative component of the heavy quark energy loss becomes necessary. A third force $\vec{f}_g = -d\vec{p}_g/dt$ hence is introduced to account for the recoil force that is experienced by the heavy quarks when they emit bremsstrahlung gluons, with $\vec{p}_g$ being the emitted gluon momentum. A higher twist calculation for the medium-induced gluon radiation is adopted from~\cite{Guo:2000nz,Zhang:2003wk}:
\begin{equation}
\frac{dN_g}{dxdk^2_{\perp}dt} = \frac{2\alpha_s P(x) \hat{q}_g}{\pi k_{\perp}^4} \sin^2\left(\frac{t-t_i}{2\tau_f}\right) \left(\frac{k_{\perp}^2}{k_{\perp}^2 + x^2M^2}\right)^4.
\end{equation}
where $x$ is the fractional energy carried by the emitted gluon, $k_{\perp}$ is the gluon transverse momentum, $P(x)$ is the splitting function and $\tau_f$ is the gluon formation time. We relate the gluon transport coefficient $\hat{q}_g$ and heavy quark transport coefficient $\hat{q}$ via the color factors $\hat{q}_g = C_A/C_F \hat{q}$ ($C_F=3, C_A=4/3$). Under this construction, the drag force, the thermal random force and the recoil force are dependent on the heavy quark transport coefficient $\hat{q}$, which characterizes the interaction strength between heavy quarks and the medium. In the literature, the spatial diffusion coefficient $D_s = 4 T^2/\hat{q}$ is more often used in the diffusion equation, and this will be the physical property that we are trying to extract from our analysis.

The evolution of the QGP medium is simulated by a (2+1)-dimensional event-by-event viscous hydrodynamical model VISHNEW~\cite{Song:2007ux,Shen:2014vra,Heinz:2015arc}, which has been updated to include both the shear and bulk viscosities with the shear-bulk coupling. The shear and bulk viscosities have been parametrized as temperature dependent. 
In our current study, the parameters related to properties of the soft medium as well as the initial condition are calibrated through an independent Bayesian analysis of light hadrons~\cite{Bernhard:2016tnd}. 

\subsubsection{Hadronization} 

When the temperature of the QGP medium drops below the critical temperature ($T_{\mathrm{c}}=154$ MeV), the medium undergoes a transition from a deconfined fluid to a confined hadron gas. The phenomenon of confinement involves non-perturbative processes and is not well understood. One often utilizes an instantaneous hadronization model to convert the fluid medium into hadrons. On the transition hypersurface, an ensemble of hadrons is generated by sampling the momentum distribution from the Cooper-Frye formula~\cite{Cooper:1974mv,Qiu:2013wca}:
\begin{equation}
E\frac{dN_i}{d^3 p} = \int_{\Sigma} f_i(x,y) p^{\mu} d^3 \sigma_{\mu}.
\end{equation}
Heavy quarks hadronize into heavy mesons within a hybrid model of instantaneous recombination and fragmentation. The momentum spectra of the meson produced by recombination is determined by the Wigner function~\cite{Cao:2013ita,Cao:2015hia}:
\begin{equation}
\frac{dN_M}{d^3p_M} = \int d^3 p_1 d^3 p_2 \frac{dN_Q}{d^3p_Q} \frac{dN_q}{d^3 p_q} f^W_M(\vec{p}_Q, \vec{p}_q) \delta(\vec{p}_M - \vec{p}_Q - \vec{p}_q).
\end{equation}
where $\vec{p}_Q$ and $\vec{p}_q$ are the momentum of heavy quark and light quark that constitute the heavy meson, $f^W_M(\vec{p}_Q, \vec{p}_q)$ is the Wigner function in terms of the overlap between the two initial partons and the final meson. A simple quantum harmonic oscillator is used to approximate the wave-function. For heavy quarks that do not recombine with light partons, a fragmentation process via PYTHIA  takes place. It is found that high momentum heavy quarks tend to fragment while lower momentum heavy quarks tend to recombine with the thermal light partons and hadronize into hadrons~\cite{Cao:2015hia}.

\subsubsection{Hadronic re-scattering}

After hadronization, the system continues expanding as an interacting hadron gas. Subsequent interactions between heavy mesons and light hadrons (e.g scattering and decay) after hadronization are modeled with UrQMD, which solves the Boltzmann equation for all the particles in the system~\cite{Bass:1998ca,Bleicher:1999xi}. UrQMD continues to evolve the system until the hadron gas is so dilute that all interactions have ceased and the system reaches its kinematic freeze-out. The particle information is then collected to calculate the final state observables that can be compared with experimental data.

In our study of D mesons, the two main observables are: the nuclear modification factor $R_{\mathrm{AA}}$, which quantifies the heavy quark in-medium energy loss and is obtained by taking the ratio of the heavy meson $p_\mathrm{T}$ spectra measured in nucleus-nucleus collisions and the reference spectra in proton-proton collisions, scaled by the binary collision number; the harmonic flow coefficients $v_n$, which are the $n^{\mathrm{th}}$ order coefficients in the azimuthal angle Fourier expansion of the emitted hadron spectra. In our calculation, the second order harmonic elliptic flow $v_2$ is calculated via both: the event-plane method~\cite{Poskanzer:1998yz} and the cumulant method~\cite{Bilandzic:2010jr}, while the triangle flow is calculated via the cumulant method~\cite{Bilandzic:2010jr}.

\subsection{\label{sec:params} Parameterization of the diffusion coefficient}
 
One of the goal of the heavy-ion community for the next few years is to quantitatively determine the heavy quark diffusion coefficient at sufficiently high precision. Since the diffusion coefficient is not a quantity that can be directly measured, its determination requires an interplay between both experiment and theory, meaning the values of the parameters which encode the heavy quark diffusion coefficient are obtained from a comparison between experimental measurement and the corresponding theoretical calculations. 

\begin{figure}[t]
\includegraphics[width=0.5\textwidth]{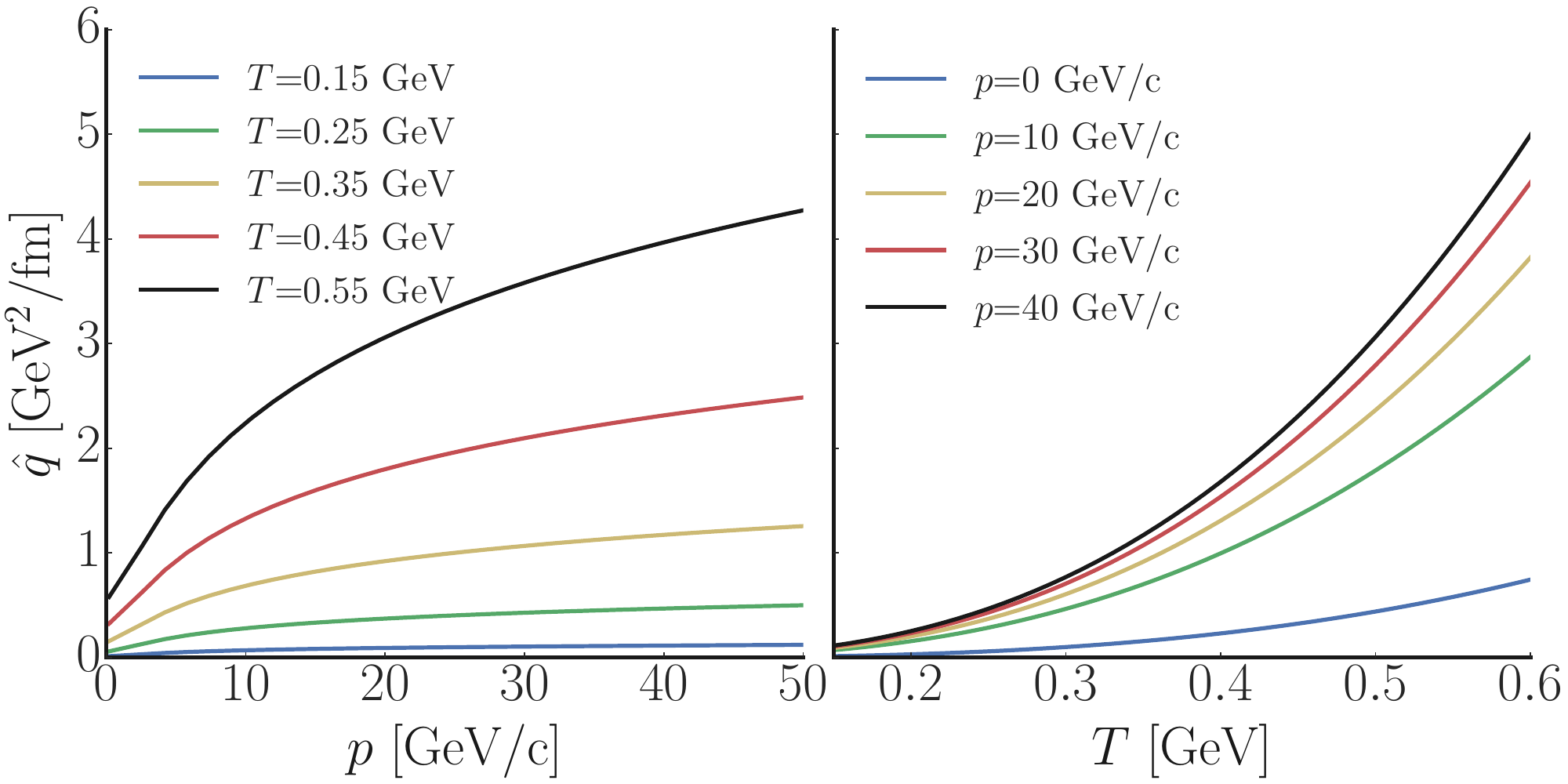}
\caption{\label{fig:qhat}(Color online) A leading order pQCD calculation of $\hat{q}$ with respect to temperature and momentum. For the calculation of $\hat{q}$, a fixed coupling $\alpha_s=0.3$ is used, and a Debye screening mass according to $m_D^2 = 4\pi \alpha_s T^2$. All $s, u, t$ channel contributions for $Qq\rightarrow Qq, Qg\rightarrow Qg$ are included.}
\end{figure}

At high temperature and large momentum, the diffusion coefficient can be calculated using perturbative QCD~\cite{Cutler:1977qm,Combridge:1978kx}: The simplest possible diagram for heavy quarks interacting with light partons is given by two $2\rightarrow 2$ elastic scattering processes ($Qq\rightarrow Qq$, $Qg\rightarrow Qg$), where the heavy quark transport coefficient equals to~\cite{Baier:1996sk}:
\begin{equation}
\hat{q} = \left<(\vec{p}_{Q'})^2 - (\hat{p}_Q \cdot \vec{p}_{Q'})^2\right>.
\end{equation}
where $\vec{p}_{Q}(\vec{p}_{Q'})$ is the in-(out-)going momentum of the heavy quark.
$\left<X\right>$ is defined as:
\begin{equation}
\begin{split}
\left<X\right> = & \frac{\gamma}{2E_Q} \int \frac{d^3 p_q}{(2\pi)^3 2 E_q}\frac{d^3 p_{Q'}}{(2\pi)^3 2 E_{Q'}} \frac{d^3 p_{q'}}{(2\pi)^3 2 E_{q'}} \cdot X \cdot f_q(\vec{p}_q) \\& (2\pi)^4 \delta^4(\vec{p}_Q + \vec{p}_q - \vec{p}_{Q'} -\vec{p}_{q'}) \sum |\mathcal{M}|^2_{Qq\rightarrow Q'q'}.
\end{split}
\end{equation}
where $\vec{p}_{q}(\vec{p}_{q'})$ is the in-(out-)going momentum of the light parton (light quark or gluon), and $\mathcal{M}_{12\rightarrow34}$ are the scattering matrices between heavy quarks and light partons. The leading-order pQCD calculation of diffusion coefficient with respect to temperature and heavy quark momentum is plotted in Fig.~\ref{fig:qhat}.

It has been found in previous comparisons to data that the perturbative calculations are not sufficient to explain the experimental findings such as the single non-photonic electron suppression stemming from decay of the heavy flavor mesons (``single electron puzzle")~\cite{Gossiaux:2010yx}, or fail to to simultaneously describe both the heavy quark nuclear modification factor $R_{\mathrm{AA}}$ and elliptic flow $v_2$ (``heavy quark $R_{\mathrm{AA}}$ and $v_2$ puzzle")~\cite{Das:2015ana}. Moreover, it has been argued that the convergence of the perturbative terms is rather poor~\cite{CaronHuot:2007gq,CaronHuot:2008uh}. In order to compensate for non-perturbative effects, one may introduce a $K$-factor to scale the scattering cross section by an ad-hoc parameter, which will be adjusted until the model is able to describe the experimental data. In this study, we use a more generalized parametrization of the diffusion coefficient, which combines a linear temperature dependent component and a pQCD component:
\begin{equation}\label{eqn:D2piT}
\begin{split}
D_s2\pi T(T, \bm{p}) & = \frac{1}{1 + (\gamma^2 p)^2} (D_s2\pi T) ^{\mathrm{linear}} \\ & + \frac{(\gamma^2 p)^2}{1 + (\gamma^2 p)^2} (D_s2\pi T)^{\mathrm{pQCD}}.
\end{split}
\end{equation}

The linear component $(D_s2\pi T)^{\mathrm{linear}}$ = $\alpha \cdot (1 + \beta (T/T_{\mathrm{c}}- 1))$, which accounts for non-perturbative effects, is the diffusion coefficient in the $p=0$ GeV/c limit and can be compared to lattice QCD calculation of the spatial diffusion coefficient at zero momentum. The pQCD component is the contribution from perturbative processes, and is related to $\hat{q}^{\mathrm{pQCD}}$, which has been calculated above, by $(D_s2\pi T)^{\mathrm{pQCD}} = 8\pi / (\hat{q}/T^3)$. It should be noted that the standard spatial diffusion coefficient $D_s$ is defined at the zero momentum $p=0$ GeV/c limit. However, in this work we use the notation $D_s$ to refer to the diffusion coefficient in the full momentum range. 

\begin{figure}[t]
\includegraphics[width=0.5\textwidth]{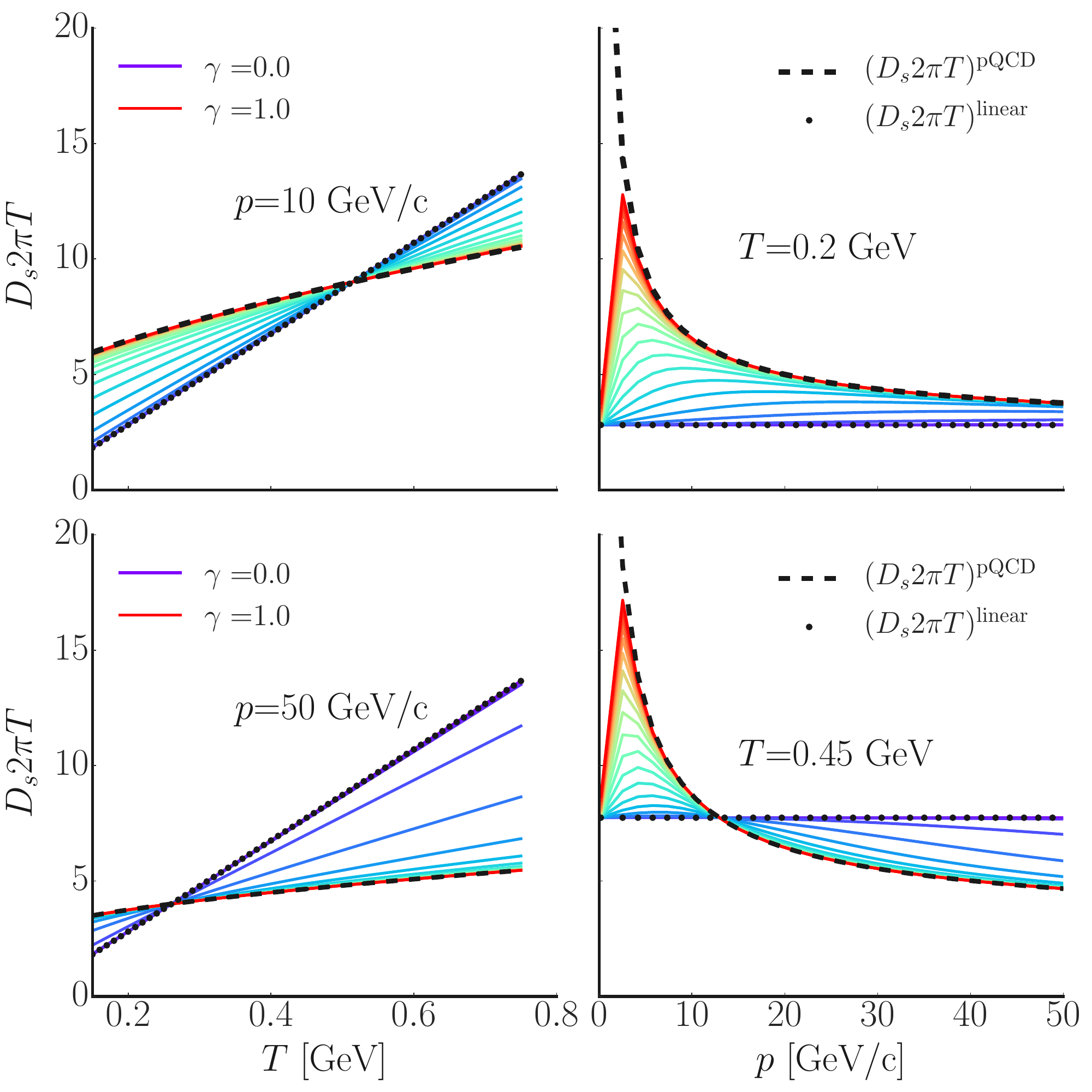}
\caption{\label{fig:gamma} (Color online) An example of the spatial diffusion coefficient parametrization. The linear component uses $(D2\pi T)^{\mathrm{linear}}=\alpha \cdot (1+\beta(T/T_{\mathrm{c}} -1))$ with $(\alpha,\beta)=(1.9,1.6)$ and is plotted as black dots. The dashed black line is the pQCD component, while the rainbow lines represent the diffusion coefficient Eqn.(\ref{eqn:D2piT}) with parameter $\gamma$ varying from 0 to 1 while the color change from violet to red.}
\end{figure} 

The parameter $\alpha$ represents the spatial diffusion coefficient at zero momentum near $T_{\mathrm{c}}$, parameter $\beta$ is the slope of $D_s2\pi T(p=0)$ above $T_{\mathrm{c}}$. The linear shape of the parametrization is inspired by the approximately linear temperature dependence of the specific shear viscosity~\cite{Bernhard:2015hxa}, as we assume an underlying relationship between the transport properties of the QGP medium~\cite{Rapp:2008qc}. The parameter $\gamma$ controls the ratio between the linear component and the pQCD component. For $p<1/\gamma^2$ the linear component dominates while for $p>1/\gamma^2$ the pQCD component is dominant. A small value of $\gamma$ indicates non-perturbative processes affect the heavy quark dynamics into the very high momentum region, and a large value of $\gamma$ indicates a quick conversion to the pQCD dominated region. To better illustrate the dependence of the spatial diffusion coefficient on $\gamma$, we plot $D_s2\pi T$ as a function of temperature (at fixed momentum) and momentum (at fixed temperature) for different values of $\gamma$ in Fig.~\ref{fig:gamma}. As shown in Fig.~\ref{fig:gamma}, the value of $\gamma$ changes from 0 to 1 while the color changes from violet to red in the reverse rainbow color scheme. For a large value of $\gamma$ (red) the combined diffusion coefficient quickly converges to the pQCD calculation, while for a small value of $\gamma$ (violet) the diffusion coefficient still follows the linear contribution even for large momenta.

\section{\label{sec:Bayesian}Parameter Calibration}
In this section we summarize the workflow of the Bayesian analysis that allows us to determine the high likelihood parameter ranges of ($\alpha, \beta, \gamma$) that govern the diffusion coefficient. More details on the Bayesian analysis can be found in ~\cite{Novak:2013bqa,Pratt:2015zsa,Bernhard:2015hxa,Bernhard:2016tnd}. 

We first evaluate the improved Langevin model for a limited number of parameter values that are selected using a Latin hypercube algorithm, and calculate the heavy quark observables as outputs from the model for these values. The mapping from inputs to outputs can then be used to train a set of Gaussian process emulators, which act as fast surrogate model of our Langevin model and are able to predict the output for any arbitrary input point in the parameter space. A Markov chain Monte Carlo (MCMC) random walk through the parameter space is then performed in order to calibrate the model parameters on the experimental data. After the MCMC equilibrates, we obtain the posterior distributions of the input parameters, and thus the posterior estimate of the parametrized diffusion coefficient.

\subsection{\label{sec:prior}Training data preparation}
\begin{table*}
\caption{\label{tab:obs} $D$-meson variables to be compared between model calculation and experimental measurements}
\begin{ruledtabular}
\begin{tabular}{ccccc}
Experiment & variables & kinematic cut & centrality & ref \\ \hline
AuAu@200 GeV & $R_{\mathrm{AA}}(p_\mathrm{T})$ 	& 6 $p_\mathrm{T}$ bins from $2-8$ GeV/c, $|y|<1$ & 0-10 & STAR~\cite{Xie:2016iwq} \\
 		     & $v_2(\mathrm{EP})(p_\mathrm{T})$	       & 8 $p_\mathrm{T}$ bins from  $1-7$ GeV/c,$|y|<1$ & 0-80  & STAR~\cite{Adamczyk:2017xur}\\
 		     & $v_2(\mathrm{EP})(p_\mathrm{T})$	       & 8 $p_\mathrm{T}$ bins from  $1-7$ GeV/c,$|y|<1$ & 10-40  \\
PbPb@2.76 TeV & $R_{\mathrm{AA}}(n_{\mathrm{{part}}})$ & 6 centrality bins, $5<p_\mathrm{T}<8$ GeV/c, $|y|<0.5$ & 0-10, 10-20,..., 40-50, 50-80 & ALICE~\cite{Adam:2015nna}\\
 				& $R_{\mathrm{AA}}(n_{\mathrm{part}})$ & 6 centrality bins, $8<p_\mathrm{T}<16$ GeV/c, $|y|<0.5$ & 0-10,
 				10-20,..., 40-50, 50-80 & \\
			  & $v_2(\mathrm{EP})(p_\mathrm{T})$ & 6 $p_\mathrm{T}$ bins from $2-16$ GeV/c, $|y|<0.8$ & 30-50 & ALICE~\cite{ALICE2017:Dmeson}\\
PbPb@5.02 TeV & $R_{\mathrm{AA}}(p_\mathrm{T})$ & 10 $p_\mathrm{T}$ bins from $3-36$ GeV/c, $|y|<0.5 $ & 30-50 & ALICE~\cite{Abelev:2014ipa} \\
			  & $v_2\{2\}(p_\mathrm{T})$ & 11 $p_\mathrm{T}$ bins from $1-40$ GeV/c, $|y|<1$ & 10-30 & CMS~\cite{CMS:2016jtu}\\
			  & $v_2\{2\}(p_\mathrm{T})$ & 8 $p_\mathrm{T}$ bins from $1-40$ GeV/c, $|y|<1$ & 30-50 &
\end{tabular}
\end{ruledtabular}
\end{table*}

Over the last few years, significant progress has been made related to the measurement of heavy flavor observables, such as yields and/or flow cumulants of heavy flavor mesons, single electrons from heavy flavor hadron semi-leptonic decays, heavy flavor tagged jets, quarkonium yields, spectra and elliptic flow etc. However, those data sets differ greatly regarding the statistical and systematic uncertainties and it is therefore not feasible to combine all of them for our current analysis. For this study, we only focus on the $D$-meson $R_{\mathrm{AA}}$ and $v_2$, which are very sensitive to the interaction mechanics between heavy quarks and the medium. These have been measured in three different systems at RHIC and the LHC: AuAu collisions at $\sqrt{s_{NN}}$ = 200 GeV, PbPb collisions at $\sqrt{s_{NN}}$ = 2.76 TeV, and PbPb collisions at $\sqrt{s_{NN}}$ = 5.02 TeV. Table~\ref{tab:obs} summarizes the measurements and kinematic/centrality cuts of the observables that have been used.

\begin{figure}[b]
\includegraphics[width=0.5\textwidth]{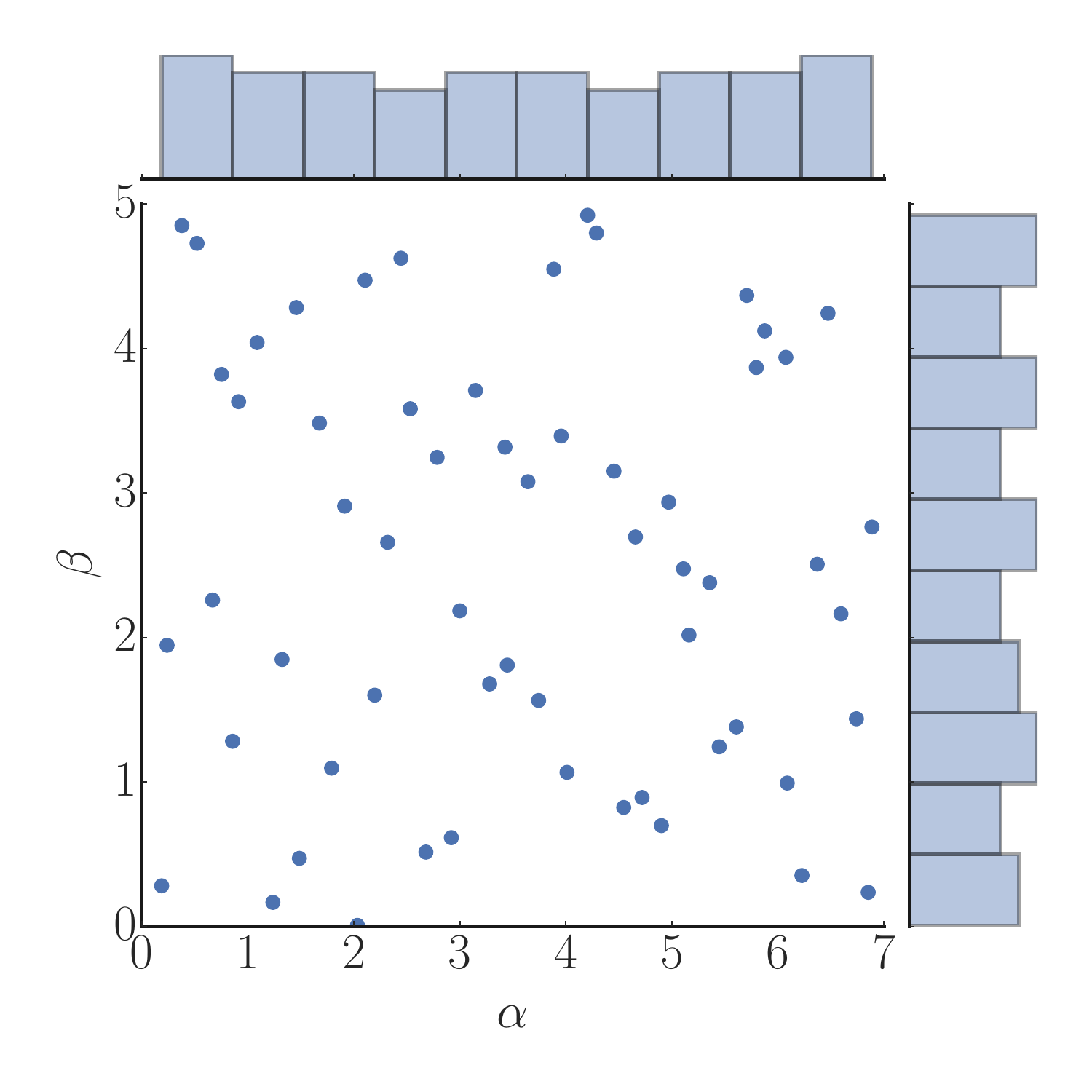}
\caption{\label{fig:latin} (Color online) 60 design points $\tilde{X}$ generated by the Latin hypercube algorithm, projected in the $(\alpha,\beta)$ dimensions. The flat histograms on the edge show an uniform prior distribution of the parameters.}
\end{figure}

\begin{figure}[t]
\includegraphics[width=0.48\textwidth]{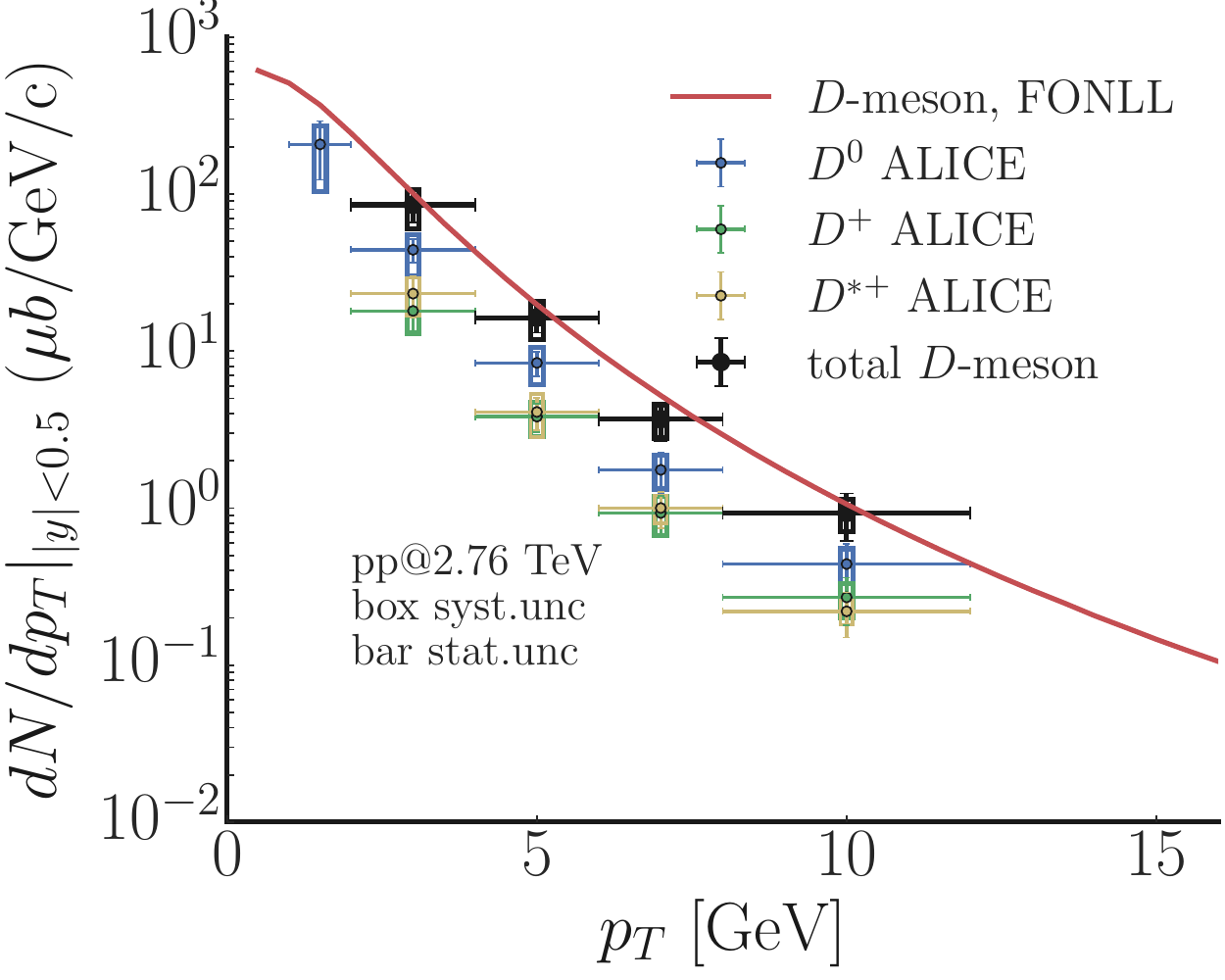}
\caption{\label{fig:pp} (Color online) Proton-proton reference spectrum calculated by FONLL+PYTHIA, used as the reference spectra in order to calculate $D$-meson $R_{\mathrm{AA}}$. The experimental results denoted {\em total D-meson} are the sum of $D^0$,$D^+$ and $D^{*+}$ data from ALICE collaboration.}
\end{figure}

\begin{figure*}[t]
\includegraphics[width=1.0\textwidth]{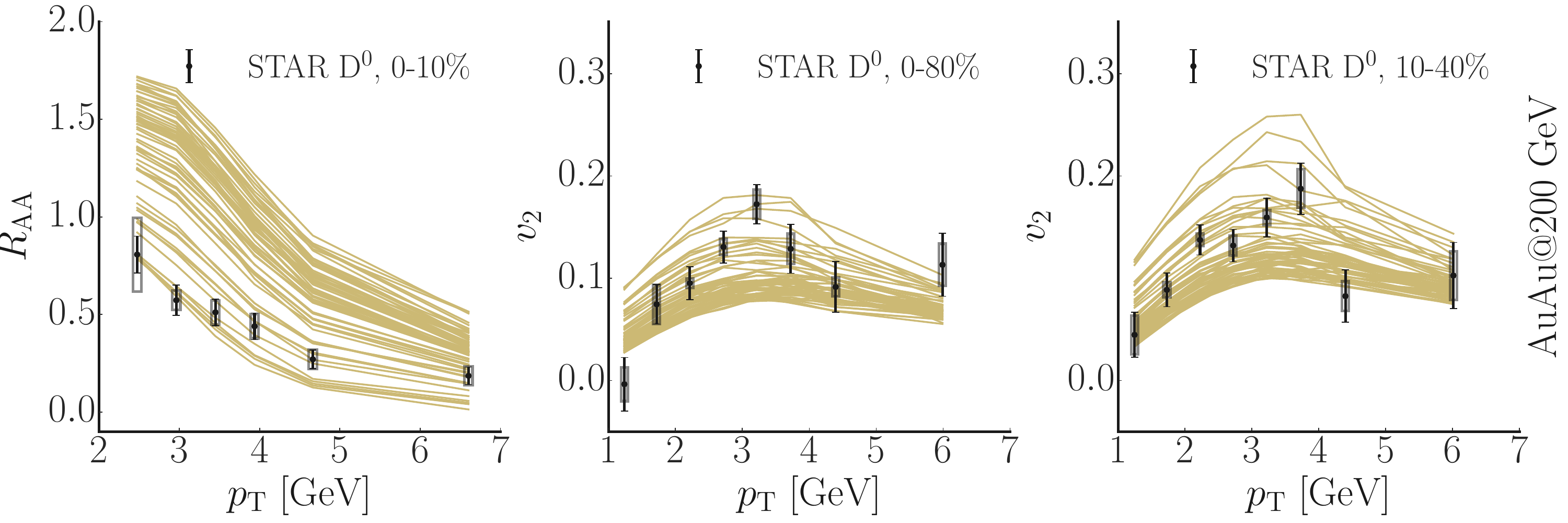}
\includegraphics[width=1.0\textwidth]{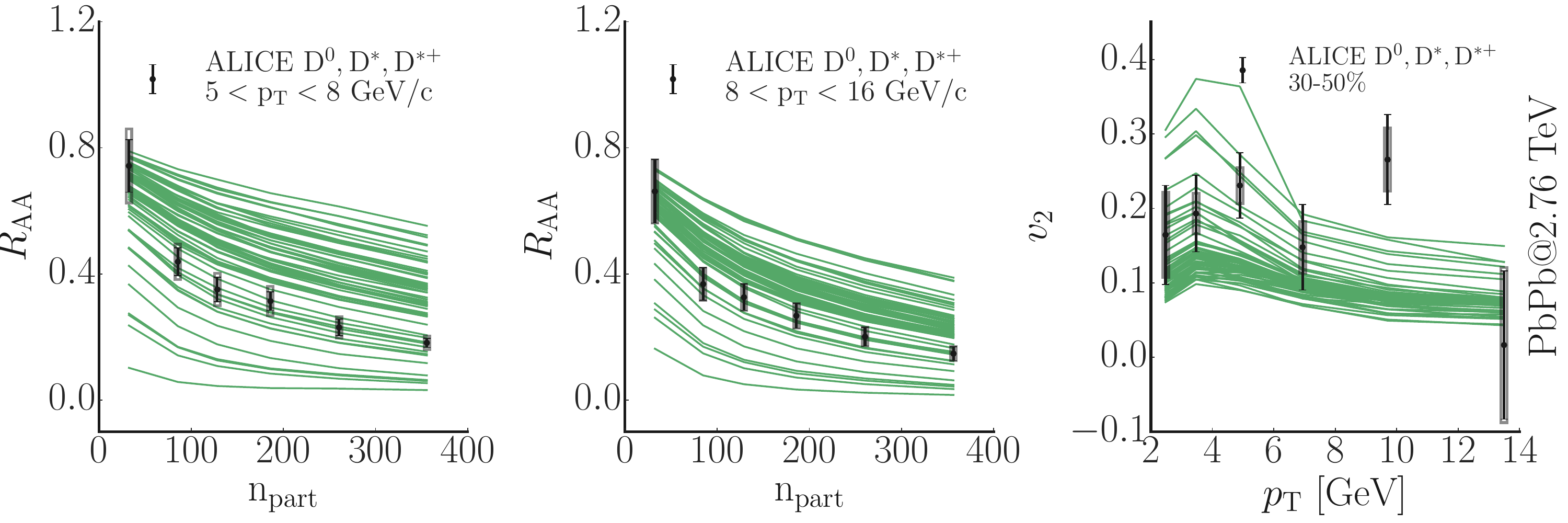}
\includegraphics[width=1.0\textwidth]{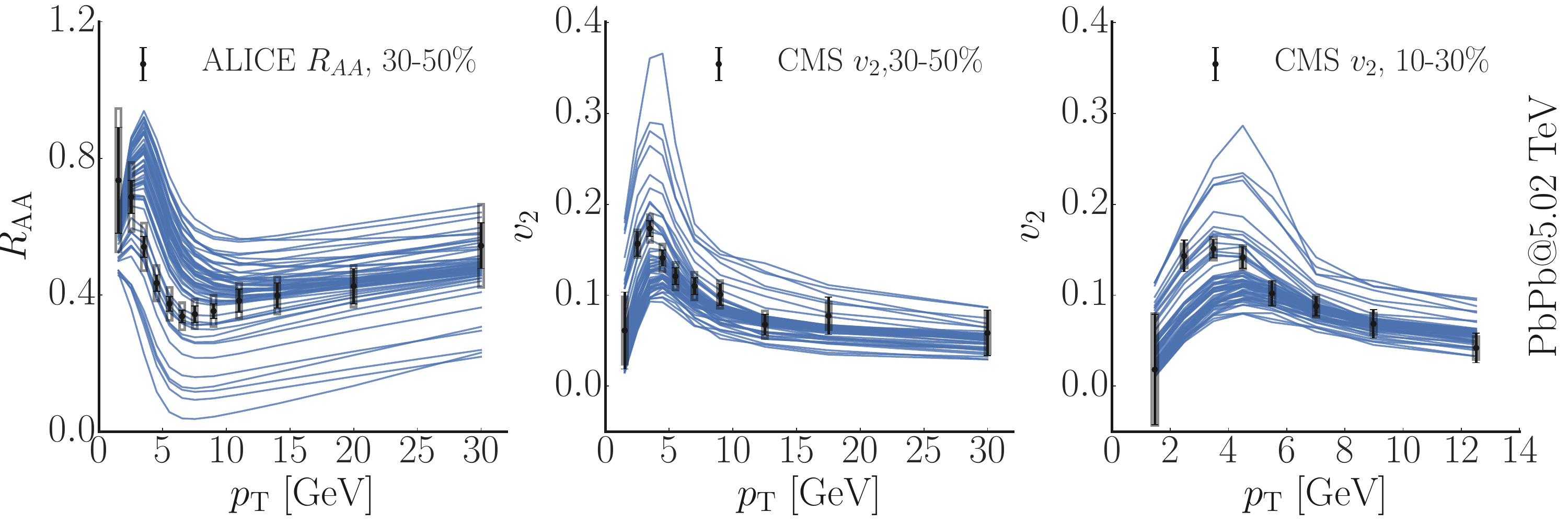}
\caption{\label{fig:prior}(Color online) Improved Langevin model calculation of $D$-meson observables, compared to experimental data spanning the full range of the explored parameter space (i.e. the ``prior"). Each frame contains 60 lines, corresponding to the 60 design points of the analysis. From top to bottom: (\textbf{top}) AuAu collisions at 200 GeV: $D$-meson $R_{\mathrm{AA}}(p_{\mathrm{T}})$ in the 0-10\% centrality bin and $v_2(p_{\mathrm{T}})$ in the 0-80\%, 10-40\% centrality bins; (\textbf{middle}) PbPb collisions at 2.76 TeV: $D$-meson $R_{\mathrm{AA}}$ as function of centrality at high momentum range $5<p_{\mathrm{T}}<8$ GeV/c and $8<p_{\mathrm{T}}<16$ GeV/c, $v_2$ as function of $p_{\mathrm{T}}$ in 30-50\% centrality bin; (\textbf{bottom}) PbPb collisions at 5.02 TeV: $D$-meson $R_{\mathrm{AA}}(p_{\mathrm{T}})$ in 30-50\% centrality bin and $v_2(p_{\mathrm{T}})$ in 30-50\% and 10-30\% centrality bins. Experimental data are measured by STAR~\cite{Xie:2016iwq,Adamczyk:2017xur}, ALCIE~\cite{Adam:2015nna,Abelev:2014ipa,ALICE2017:Dmeson} and CMS~\cite{CMS:2016jtu}.}
\end{figure*}

To illustrate the degree to which the model's calculation are affected by the particular form of the temperature and momentum dependence of the diffusion coefficient, we first sample 60 design points ($\tilde{X} = (\mathbf{x}_1, \cdots, \mathbf{x}_{60})^T, \mathbf{x}_i=(\alpha, \beta,\gamma)^T$) in the parameter space (the tilde symbol $\tilde{x},\tilde{y}$ represents the training datasets, whose outputs are calculated from the physical model, it is used to distinguish the predicted datasets, which are later obtained from emulators and labeled with the star symbol $x_*,y_*$). Those design points are semi-randomly sampled in the 3-dimensional parameter space using a Latin hypercube algorithm, which aims at spreading the samples evenly across all possible values~\cite{Morris:1995}, and therefore, a small amount of samples $\cal{O}\mathrm{(10p)}$ is sufficient to train the Gaussian process emulators to interpolate the $p$-dimensional parameter space. The parameter space is deliberately chosen to be wide enough in order to cover the full range of likely values, and is listed in Tab.~\ref{tab:range}. Figure~\ref{fig:latin} visualizes the uniform distribution the initial design points, projected in the $(\alpha, \beta)$ plane.

\begin{table}
\caption{\label{tab:range} Prior range and description for the parameters that determines the diffusion coefficients}
\begin{ruledtabular}
\begin{tabular}{ccc}
Parameter & Description & Range\\
\hline
$\alpha$ & $D_s2\pi T$ at $T_{\mathrm{c}}$ & 0.1-7.0\\

$\beta$ & slope of $(D_s2\pi T)^{\mathrm{linear}}$above $T_{\mathrm{c}}$ & 0-5.0\\
$\gamma$ & ratio between $D_s^{\mathrm{linear}}$ and$D_s^{\mathrm{pQCD}}$ & 0.0-0.6\\
\end{tabular}
\end{ruledtabular}
\end{table}

At each of the design points, we generate 5000 minimum bias hydro events and for each hydro event, heavy quarks are oversampled to reduce the theoretical statistical uncertainty. The heavy meson observables are then calculated as the following: the events are first binned into different centrality classes according to the final state charged hadron multiplicity $N_{ch}$ at mid-rapidity. The $D$-meson selection is based on the corresponding experimental kinematic cuts. In order to calculate the nuclear modification factor $R_{\mathrm{AA}}$, a proton-proton collision reference is needed. It is calculated using a heavy quark FONLL distribution followed by a fragmentation process performed by PYTHIA, and the $D$-meson yields in the proton-proton reference are compared with experimental data in Fig.~\ref{fig:pp}. For the calculation of $D$-meson elliptic flow, we try to match the experimental methods as far as possible, therefore for the AuAu collision and PbPb collision at 2.76 TeV, an event-plane (approximated by the initial participant-plane) method is used, while for the PbPb collision at 5.02 TeV, the two-particle cumulant method is used, though little difference has been noticed for the two different methods~\cite{Abelev:2014ipa,Adamczyk:2017xur}. In total there are 69 experimental data points to calibrate against.

Figure~\ref{fig:prior} compares the 60 sets of model calculation with the experimental data (black dots with errorbars). 
We can see that the model's outputs span a wide range in observable space as the input parameters have been randomly distributed in the parameter space. These input $\tilde{X}_{n\times p}$ and output  $\tilde{Y}_{n \times m}$ matrices (where $n=60$ is the number of input parameter points, $p=3$ is the dimension of input parameters, $m=69$ is the dimension of output at each of the input point)  will then be used to train the Gaussian process emulators.

\subsection{\label{sec:Gaussian}Gaussian process emulator}
In order to calibrate our parameters, a random walk throughout parameter space will be performed, where each step is accepted or rejected according to the relative likelihood. Taking a random walk throughout the 3-dimensional parameter space requires $\cal{O}$(1000) steps and the number increases exponentially if we try to include more parameters. At each step one needs to generate a sample of events and calculate the model's output for the observables in order to evaluate the likelihood and take action for the next step. Given the amount of computational time required to evaluate the likelihood at one point in parameter space, such a method is not feasible. To overcome this difficulty, a set of emulators is used to function as a fast surrogate model that can predict the physical model's output at any arbitrary point in the parameter space. In this study, we construct the Gaussian process (GP) emulator, which is an mapping from a n-dimensional input space to a normal distributed output. It not only interpolates and predicts the model output after being trained, but also provides the uncertainties of its prediction. More details of the GP emulator can be found in Ref.~\cite{Rasmussen:2006GP}. Here we will briefly summarize the basic idea of GP emulator.

Consider a physical model (e.g. our improved Langevin framework), whose output of a physical process is determined by a set of input parameters $y=f(\mathbf{x})$. We suppose that the physical model has been evaluated at $n$ input points in the $p$-dimensional parameter space, (the input parameter matrix $\tilde{X} = (\mathbf{x}_1, \dots,\mathbf{x}_n)^T$). At each input point, the model has one output $y_i = f(\mathbf{x}_i)$, yields to an $n$-dimensional output vector $\tilde{\mathbf{y}}$:
\begin{eqnarray}
\tilde{X} = \begin{pmatrix}
x_{11} & ... & x_{1p}\\ 
... & & ...\\
x_{n1} & ... & x_{np} 
\end{pmatrix}
\Rightarrow
\tilde{\mathbf{y}}=
\begin{pmatrix}
y_1\\... \\y_n 
\end{pmatrix}.
\end{eqnarray}

The output $\tilde{\mathbf{y}}$ can be viewed as a conditioned Gaussian process which is a collection of normal distributions:
\begin{equation}
\tilde{\mathbf{y}} = \mathcal{GP}(\tilde{X}) \sim \mathcal{N}(\mu(\tilde{X}), K_{\tilde{X}, \tilde{X}}).
\end{equation}
where $\mu(\tilde{X})$ is the mean vector of each input, and
\begin{equation}\label{eqn:cov}
K_{\tilde{X}, \tilde{X}}= 
\begin{pmatrix}
\sigma(\mathbf{x}_1,\mathbf{x}_1) &\cdots  & \sigma(\mathbf{x}_1,\mathbf{x}_n)\\ 
\vdots & \ddots & \vdots\\ 
 \sigma(\mathbf{x}_n,\mathbf{x}_1)& \cdots & \sigma(\mathbf{x}_n,\mathbf{x}_n)
\end{pmatrix}.
\end{equation}
is the covariance matrix. It is constructed by the covariance function $\sigma(\mathbf{x},\mathbf{x}')$ and characterizes the correlation between different inputs.

In order to predict the model output $y_*$ at any other input $\mathbf{x}_*$ (the star symbol are used to represent the datasets whose outputs are predicted from the emulators), one can write the joint multivariate normal distribution:
\begin{equation}
\begin{pmatrix}
y_* \\ \tilde{\mathbf{y}}
\end{pmatrix}
\sim
\mathcal{N}\left(\begin{pmatrix}
\mu(\mathbf{x}_*)\\ \mu(\tilde{X}) 
\end{pmatrix},
\begin{pmatrix}
K_{*,*} &K_{*,\tilde{X}} \\ 
K_{\tilde{X},*} & K_{\tilde{X},\tilde{X}} 
\end{pmatrix}
\right).
\end{equation}
$K_{*,*}$, $K_{*,\tilde{X}}$ and $K_{*,\tilde{X}}$ have the same form as Eqn.~(\ref{eqn:cov}) but with different $\mathbf{x}$. The distribution of a predictive output $y_*$ can be solved by:
\begin{align}
& y_* \sim \mathcal{N}(\mu, K) \\
& \mu = \mu(\mathbf{x}_*) + K_{*,\tilde{X}} K^{-1}_{\tilde{X},\tilde{X}} (\tilde{y} - \mu(\tilde{X})) \\
& K = K_{*,*} - K_{*,\tilde{X}} K^{-1}_{\tilde{X},\tilde{X}} K_{\tilde{X},*}.
\end{align}
With the knowledge of a training dataset containing a set of inputs ($\tilde{X}$) and outputs ($\tilde{y}$), and the covariance matrix ($K$), one is able to solve the equations and calculate the distribution for any other input $\mathbf{x}_*$.

The inference of the Gaussian process is determined by the covariance function $\sigma(\mathbf{x}, \mathbf{x}')$. Variance choice can be made for the covariance function, based on our knowledge and assumption of the input parameters. In this study we use a popular squared-exponential function with a noise term:
\begin{equation}
\sigma(\mathbf{x},\mathbf{x}') = \sigma^2_{\mathcal{G}} \exp \left[-\sum_{k=1}^{m} \frac{(x_k-x_k')^2}{2l_k^2}\right] + \sigma^2_n \delta_{\mathbf{x},\mathbf{x}'}.
\end{equation}
The squared-exponential covariance function implies our intuition that the inputs in the parameter space that are close to each other are highly correlated, whilst those far away are uncorrelated, the correlation strength between pairs of inputs is controlled by the hyperparameter $(\sigma_{\mathcal{G}}, l_k)$. With this covariance function, the Gaussian process is very smooth as the covariance is infinitely differentiable. In this study, the hyperparameters $(\sigma_{\mathcal{G}}, l_k, \sigma_n)$ are determined in a manner of ``best fit parameters" which maximizes the parameter likelihood function\cite{2014arXiv1403.6015A}. 

\subsection{\label{sec:PCA}Principle component analysis}
\begin{figure}[t]
\includegraphics[width=0.45\textwidth]{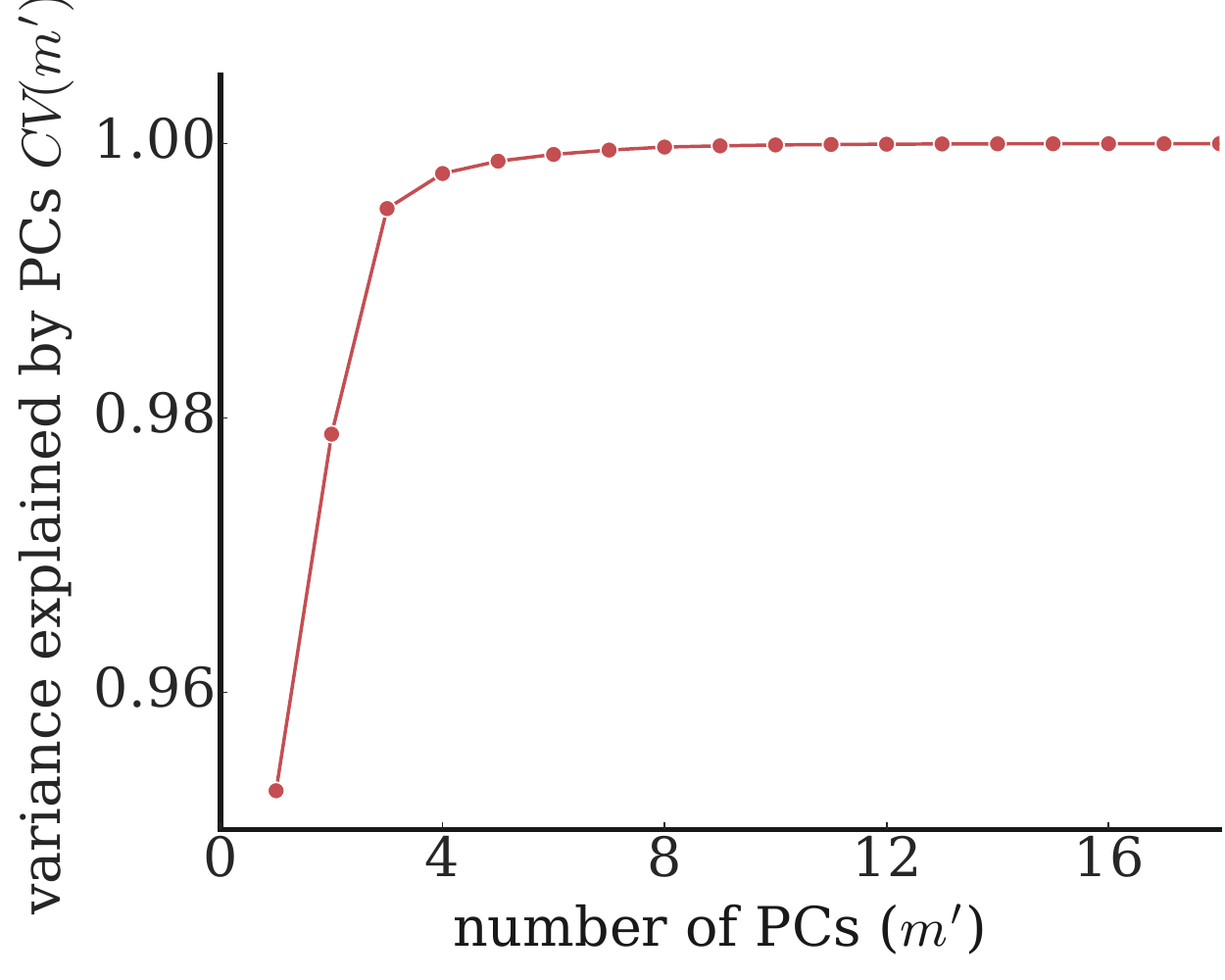}
\caption{\label{fig:variance} (Color online) Cumulative variance explained by the first $m'$ principal components. The first few PCs are able to explain most of the total variance}
\end{figure}

A Gaussian process is essentially a mapping from an input vector to a scalar output. The output from our Langevin mode is an $m$-dimensional vector ($m=69$):
\begin{eqnarray}
\tilde{X} = \begin{pmatrix}
x_{11} & ... & x_{1p}\\ 
... & & ...\\
x_{n1} & ... & x_{np} 
\end{pmatrix}
\Rightarrow
\tilde{Y}=
\begin{pmatrix}
y_{11} & ... & y_{1m}\\... \\y_{n1} & ... & y_{nm} 
\end{pmatrix}.
\end{eqnarray}

One can construct a GP emulator for each of the observables. However, as the elements in the output are highly correlated, it is useful to reduce a high dimensional and correlated output to a lower dimensional and orthogonal principal components (PCs), which are the linear combinations of the output observables and that provide information on the most important components of the dataset.

In practice, the training output $\tilde{Y}$ is standardized (by subtracting the mean and divided by the standard deviation for each observable), and decomposed via the singular value decomposition (SVD):
\begin{equation}
Y_{m\times n} = U_{m\times n} S_{n \times n} V^T_{n\times n}.
\end{equation}
The columns of $U$($V^T$) are the left(right)-singular vector of $Y$, which are sets of orthogonal eigenvectors of $YY^T$($Y^TY$). The output matrix $Y$ can then be transformed into principal component space:
\begin{equation}
Z = \sqrt{n} Y V.
\end{equation}
$S$ is the diagonal matrix whose diagonal elements $\lambda_{i,(i=1,\cdots,n)}$ are the squared roots of the eigenvalues of $Y^TY$. The eigenvalues $\lambda_i$ are proportional to the variance that contributed the $i$-th PC, and are sorted into descending order. The cumulative variance explained by the first $m'$-th PCs ($m'\leqslant m$) then equals to:
\begin{equation}
CV(m') = \frac{\sum_{i=1}^{m'}\lambda_i}{\sum_{i=1}^{m}\lambda_i}.
\end{equation}
As shown in fig.~\ref{fig:variance}, the first few PCs are sufficient to explain most of the variance of the model outputs. In this study, we use 8 PCs and for each PC a GP emulator is constructed, which is a significant reduction from the original 69 GPs that mapped directly onto the number of data points in the calibration dataset. 
\begin{figure}[t]
\includegraphics[width=0.5\textwidth]{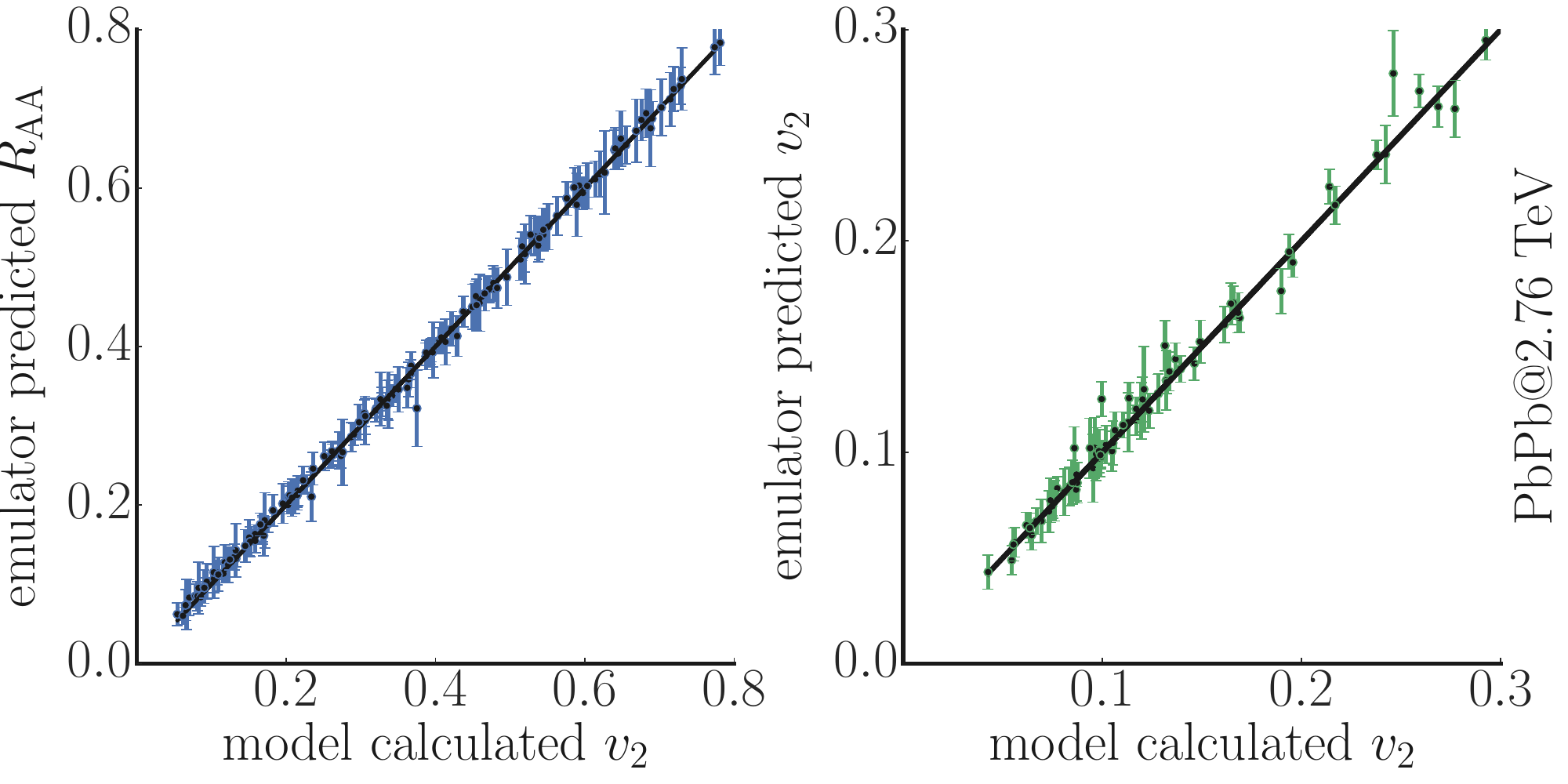}
\caption{\label{fig:validation} (Color online) Validation of the Gaussian emulators for PbPb collisions at 2.76 TeV. Each point represents the emulator's predicted value with respect to the model calculated (true) value.\textbf{(left:)} prediction for $D$-meson $v_2$ at 30-50\%; \textbf{(right:)} prediction for high momentum $D$-meson $R_{\mathrm{AA}}$ at different centrality bins.}
\end{figure}

\begin{figure*}
\includegraphics[width=1.0\textwidth]{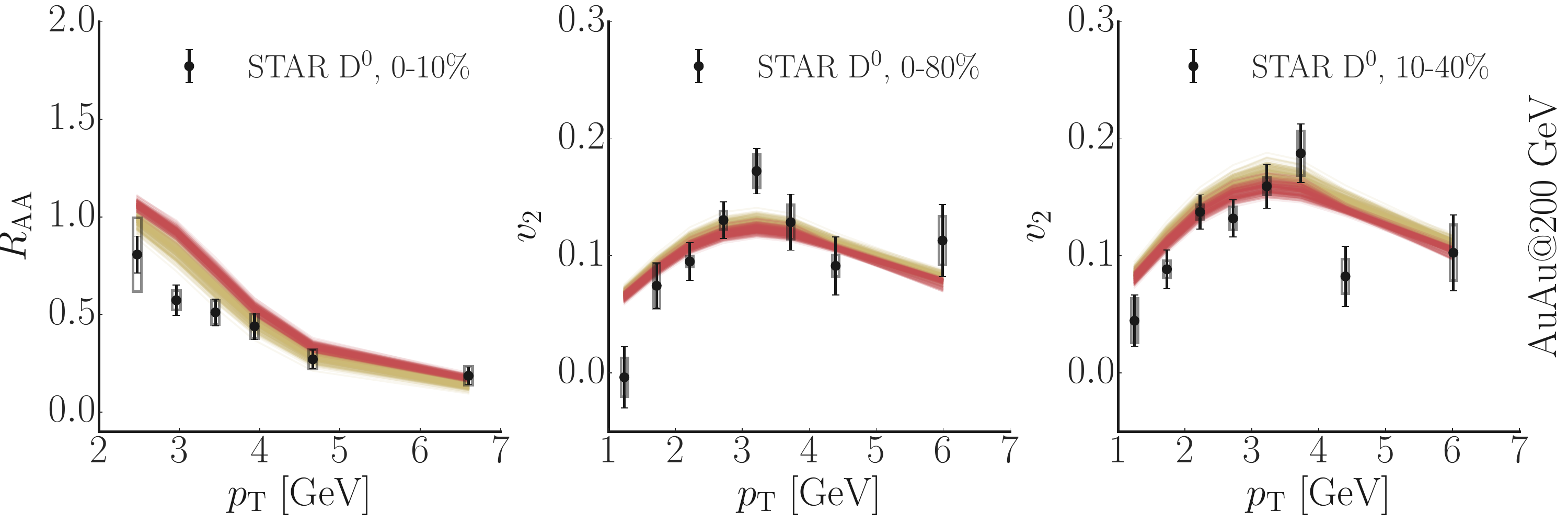}
\includegraphics[width=1.0\textwidth]{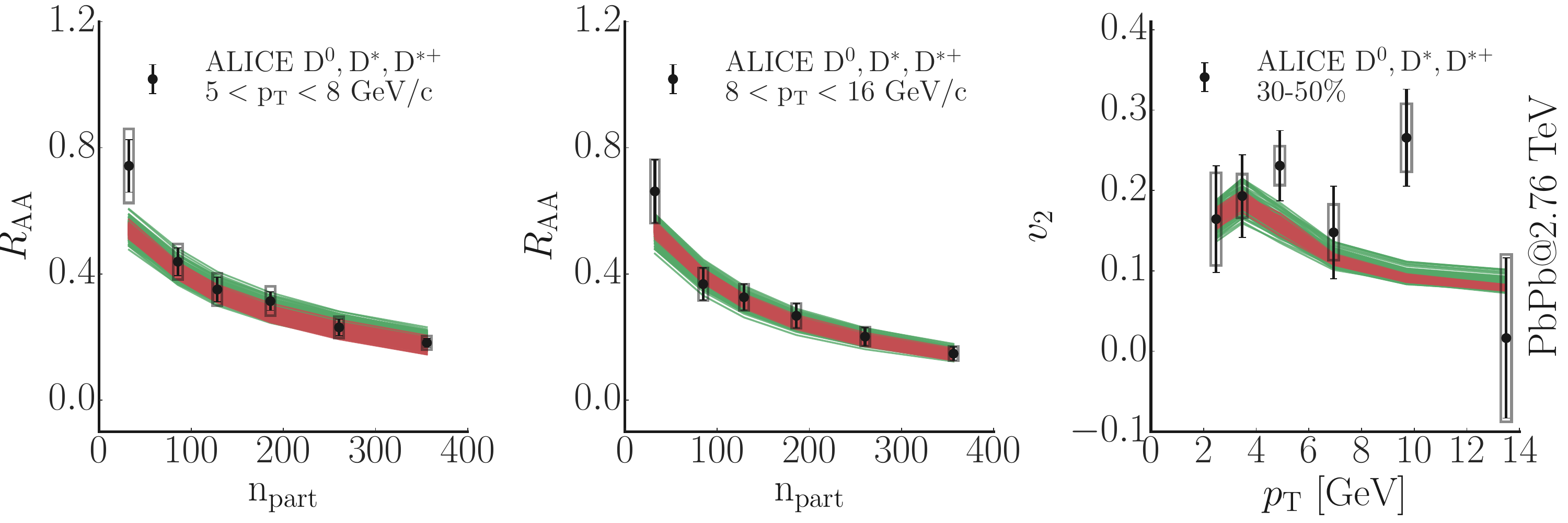}
\includegraphics[width=1.0\textwidth]{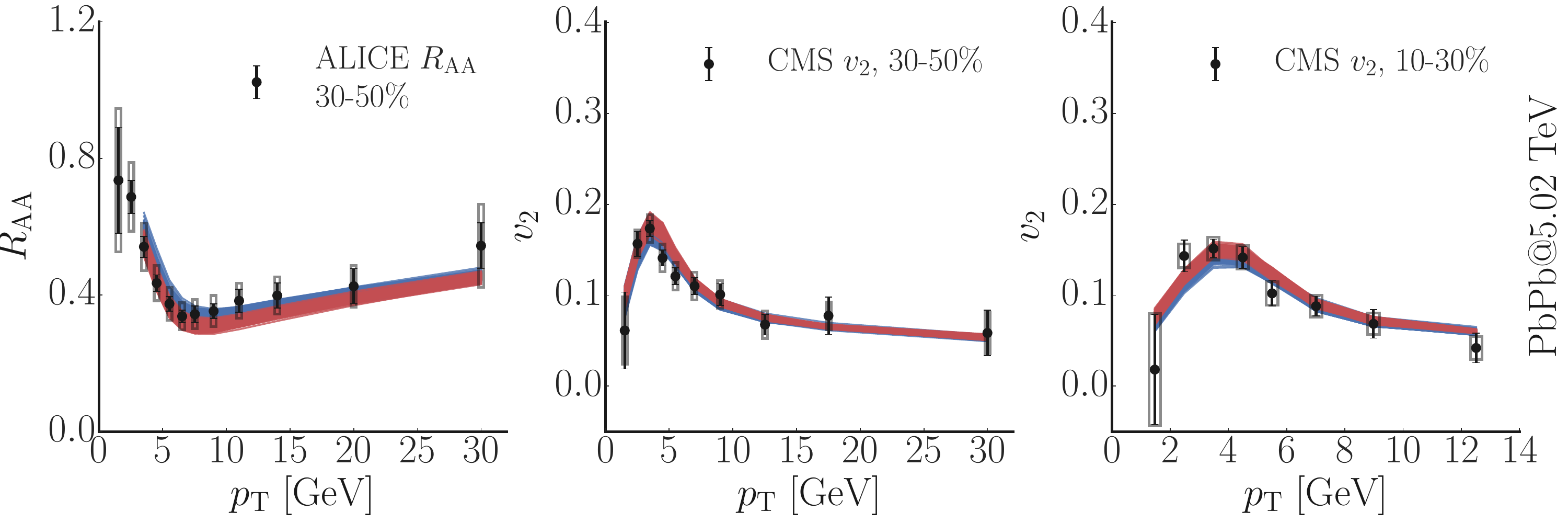}           
\caption{\label{fig:posterior}(Color online) Emulator predictions for 200 random input parameters sampled from the posterior distributions. This figure is similar to Fig.\ref{fig:prior} but with the input parameters chosen from the posterior distribution, and the outputs are predictions from the GP emulators.}
\end{figure*}

Once the principal component $\mathbf{z}$ has been determined, we obtain the outputs in the physical observable space by the inverse transformation:
\begin{equation}
\mathbf{y} = \frac{1}{\sqrt{n}} \mathbf{z}  V.
\end{equation}

In order to test the emulators' ability to predict the models' output, we generate another 15 sets of test inputs and perform the full Langevin model calculation at each of these test inputs. Meanwhile the trained GP emulators also predict the output for each of these input parameters. Figure~\ref{fig:validation} compares the prediction from the emulators to the calculation from the improved Langevin model for PbPb collisions at 2.76 TeV. For each test input point, 18 observation of $R_{\mathrm{AA}}$ and $v_2$ are calculated at different centralities and $p_{\mathrm{T}}$ bins. The black lines are the $y=x$ reference, and each dot represent the emulators' prediction with respect to the models' calculation. As visualized in Fig.~\ref{fig:validation} the GP emulators in general work very well. We should note that the emulators provide the uncertainty for each prediction, therefore the errorbars shown in the figure correctly capture the uncertainties underlying in the emulation.

\subsection{\label{sec:MCMC}MCMC calibration}
According to Bayes theorem, the probability for the true parameter $\mathbf{x}$ given the experimental data $\mathbf{y}_{exp}$ and observed $(\tilde{X}, \tilde{Y})$ is proportional to the likelihood of the parameter $\mathbf{x}$ and its prior distribution:
\begin{equation}
P(\mathbf{x}|\tilde{X}, \tilde{Y}, \mathbf{y}_{\mathrm{exp}}) \propto P(\tilde{X}, \tilde{Y}, \mathbf{y}_{\mathrm{exp}}|\mathbf{x}) P (\mathbf{x}).
\end{equation}
where $P(\mathbf{x}|\tilde{X}, \tilde{Y}, \mathbf{y}_{\mathrm{exp}})$ is the posterior distribution of parameters $\mathbf{x}$ given the observation of $(\tilde{X},\tilde{Y},\mathbf{y}_{\mathrm{exp}})$, which is our main results from this analysis; $P(\tilde{X}, \tilde{Y}, \mathbf{y}_{\mathrm{exp}}|\mathbf{x})$ is the likelihood of observing $(\tilde{X},\tilde{Y},\mathbf{y}_{\mathrm{exp}})$ given the parameter $\mathbf{x}$, and $P (\mathbf{x})$ is the prior distribution of parameter $\mathbf{x}$. 

Our goal here is to find the posterior probability distribution of the parameters $P(\mathbf{x}|\tilde{X}, \tilde{Y}, \mathbf{y}_{\mathrm{exp}})$ which would optimally reproduce the experimental data using our improved Langevin model. In order to determine the posterior distribution we perform a MCMC random walk in parameter space following the Metropolis-Hasting algorithm~\cite{ForemanMackey:2012ig}. During the random walk, each step is accepted or rejected according to the relative likelihood. Assuming a Gaussian structure for the uncertainties, the log likelihood function has the following form as a function of the output $\mathbf{y}$:
\begin{equation}
\begin{split}
\log P(\tilde{X}, \tilde{Y}, \mathbf{y}_{\mathrm{exp}}| \mathbf{x}) & = -\frac{1}{2}(\mathbf{y}-\mathbf{y}_{\mathrm{exp}})^T \Sigma^{-1} (\mathbf{y}-\mathbf{y}_{\mathrm{exp}}) \\ & - \frac{1}{2} \log| \Sigma| -\frac{m}{2}\log 2\pi .
\end{split}
\end{equation}

$\Sigma$ is the covariance matrix which accounts for all quantifiable uncertainties. There are various contributions to these uncertainties, such as the emulator prediction uncertainty, the experimental statistic and systematic uncertainties, and the physical model statistic and systematic uncertainties. Identifying all these uncertainties can be very difficult, especially for systematic uncertainties which in general may have correlations among each other. 
For the current analysis we only consider experimental statistical and uncorrelated systematic uncertainties:
\begin{equation}
\Sigma = \mathrm{diag}(\sigma^2_{\mathrm{stat}}) + \mathrm{diag}(\sigma^2_{\mathrm{sys}})  + \mathrm{diag}(\sigma^2_{\mathrm{GP}}) .
\end{equation} 
where $\sigma_{\mathrm{stat}}$ is the experimental statistical error,  $\sigma_{\mathrm{sys}}$ is the experimental systematic error (uncorrelated) for each observable, $\sigma_{\mathrm{GP}}$ is the theoretical uncertainty from Gaussian process emulator predictions. At the current stage, all the uncertainties are assumed to be uncorrelated for the purpose of simplicity as well as maximizing the overall constraint.

\section{\label{sec:results}Results}
\subsection{\label{sec:posterior}Posterior distributions}
To evaluate the success of the calibration, 200 points in parameter space are randomly chosen from the equilibrated MCMC trace and evaluated by the Gaussian emulators. Figure~\ref{fig:posterior} visualizes the corresponding observables and compares it with the experimental data. The presentation is similar to Fig.~\ref{fig:prior} but with calibrated parameters. For each plot, two posterior outputs are presented, each one corresponding to an independent Bayesian analysis but calibrated on different experimental datasets. The red lines correspond to the Bayesian analysis calibrated on all the output observables listed in Tab.~\ref{tab:obs}, whereas the yellow, green and blue lines correspond to calibrations on data of a single beam energy. We find that after calibration, our improved Langevin approach is capable of describing the experimental data reasonably well. The biggest deviations are found for a few $R_{\mathrm{AA}}$ points at very peripheral centrality and low $p_{\mathrm{T}}$: peripheral collisions are not well described by our hydrodynamical background. Also, the modeling of hadronization in the low $p_{\mathrm{T}}$ region is challenging due to significant non-perturbative effects. 
       
\begin{figure}
\includegraphics[width=0.5\textwidth]{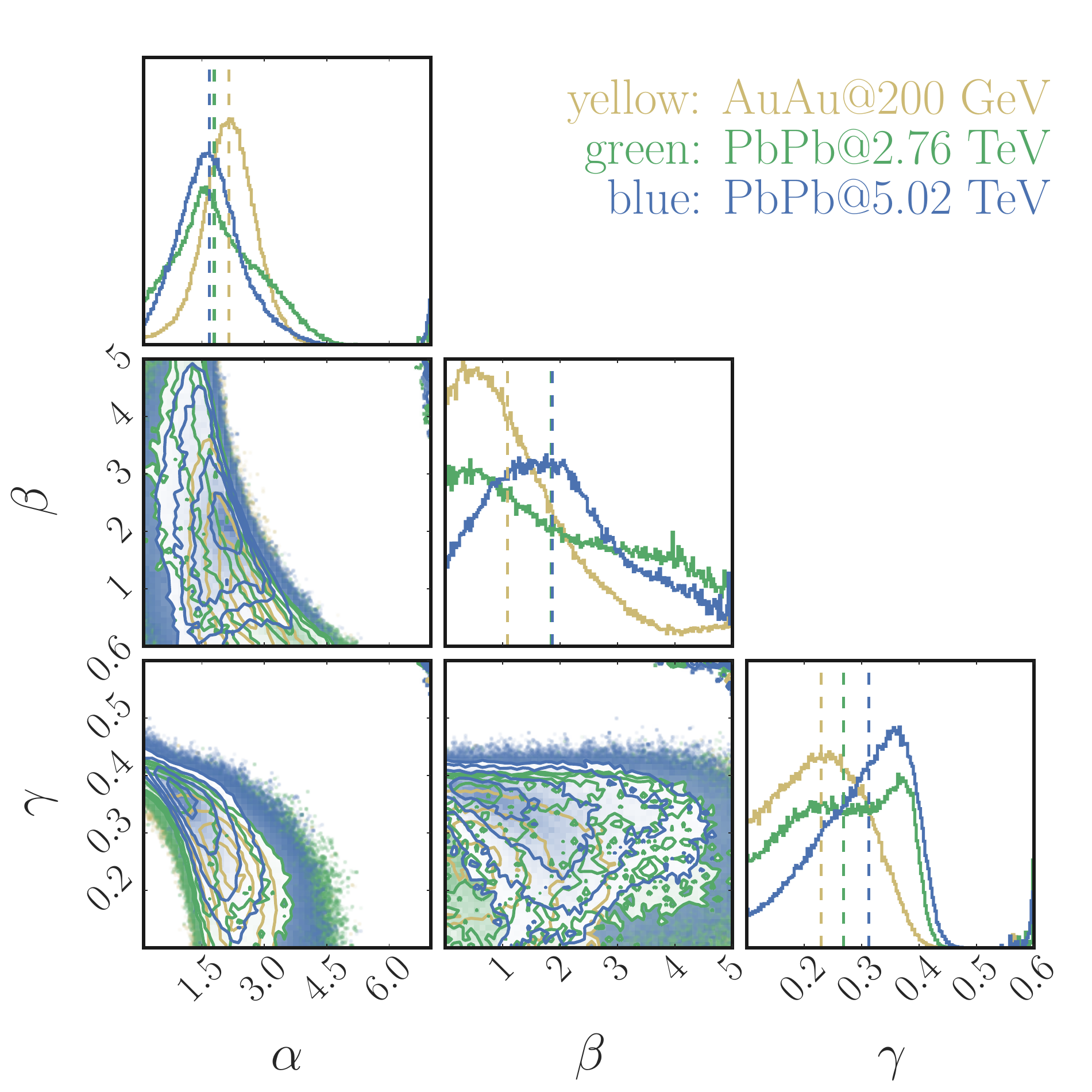}
\caption{\label{fig:corners_individual} (Color online) Posterior distributions of the diffusion coefficient parameters $(\alpha, \beta, \gamma)$ for each individual collision system. The diagonal plots are the histogram of the MCMC samples with other parameters integrated out. The off-diagonal plots display the joint distributions between pairs of parameters. The three different colors refer to the three different analyses that calibrate on three different sets of data.}
\end{figure}
\begin{figure}
\includegraphics[width=0.5\textwidth]{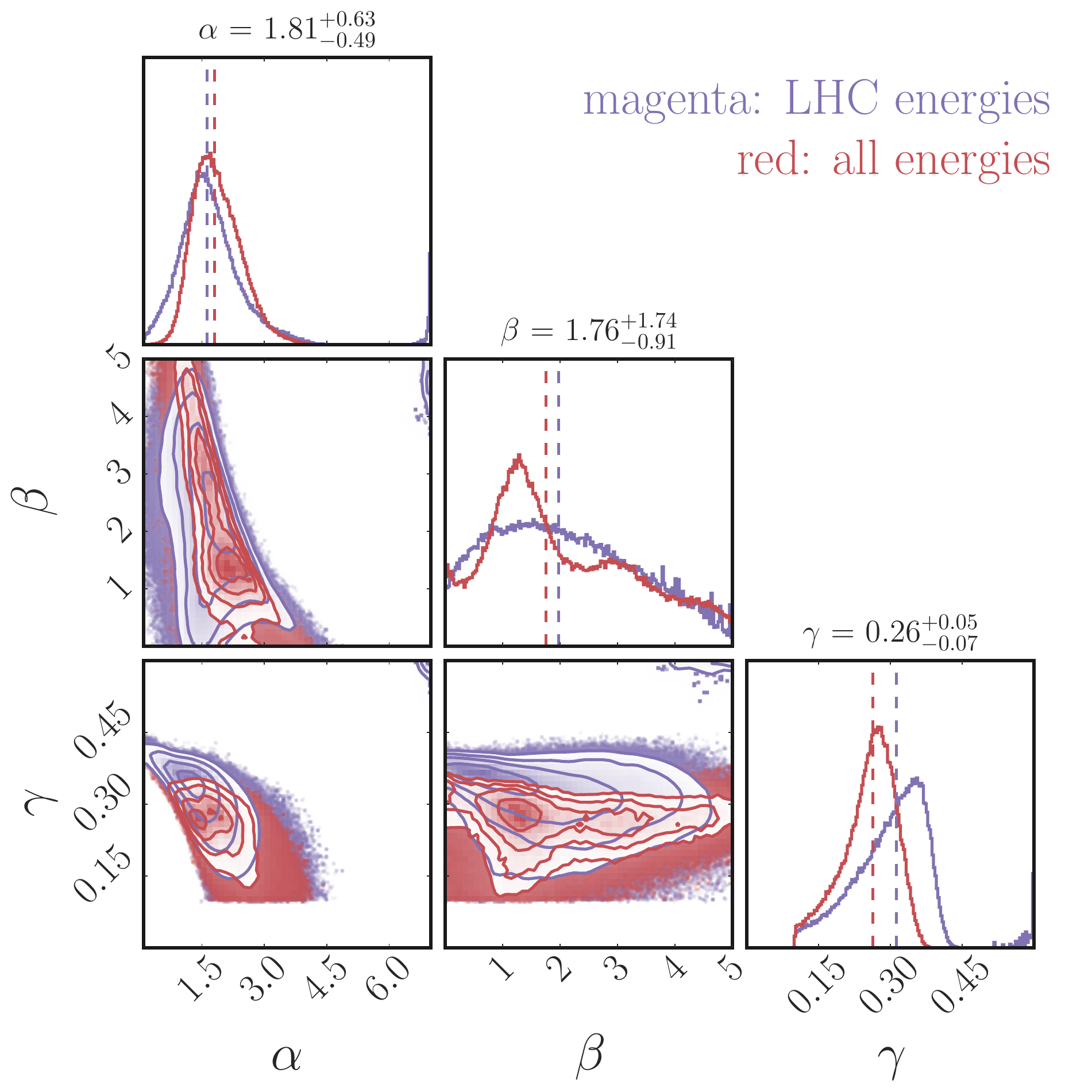}
\caption{\label{fig:corners_all} (Color online) Posterior distributions of the diffusion coefficient parameters $(\alpha, \beta, \gamma)$ for the analysis combining all three collision systems. The red color refers to the result that combines only the LHC energies: PbPb collisions at 2.76 and 5.02 TeV; the magenta color refers to the analysis combining all three energies at RHIC and LHC.}
\end{figure}

\begin{figure*}[t]
\includegraphics[width=1.0\textwidth]{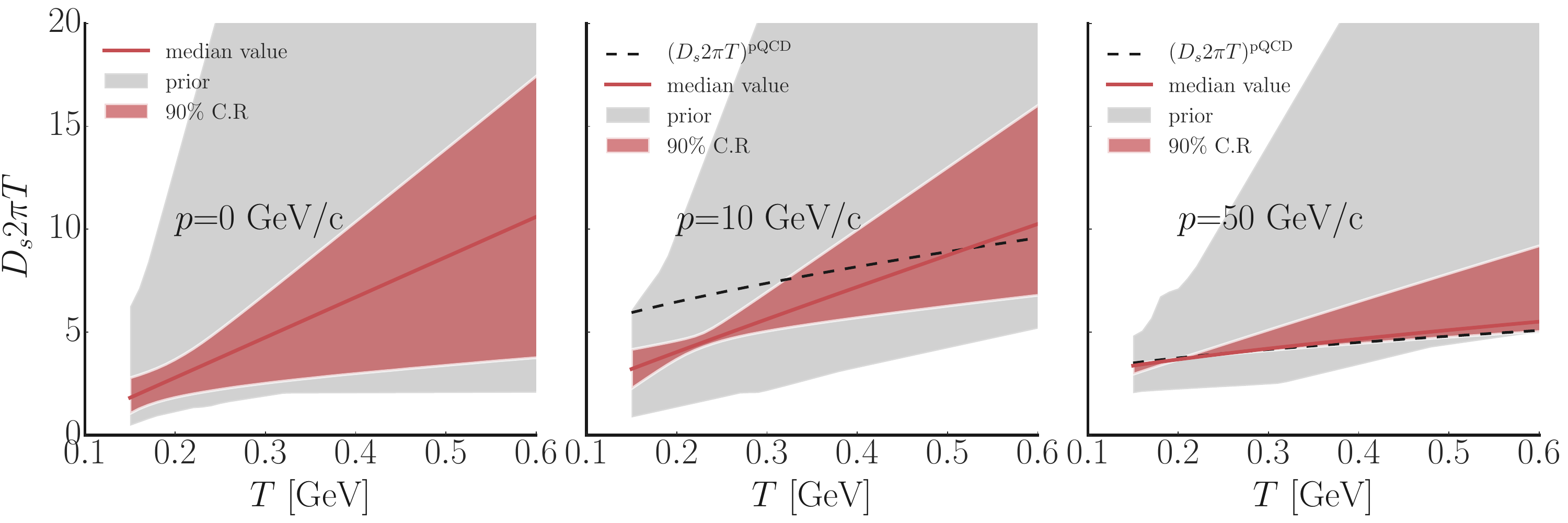}
\includegraphics[width=1.0\textwidth]{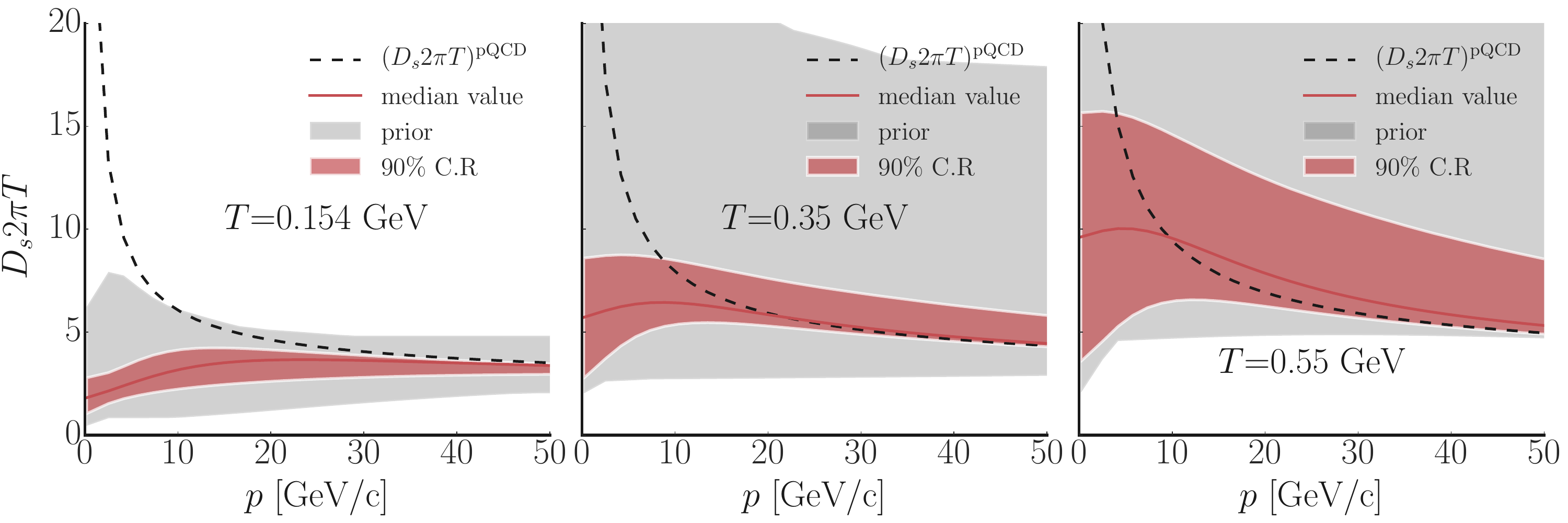}
\caption{\label{fig:D2piT}(Color online) Posterior results of the spatial diffusion coefficient from the Bayesian analysis calibrated on the combined dataset from three different systems at RHIC and the LHC.(\textbf{top}) the spatial diffusion coefficient $D_s2\pi T$ as a function of temperature at fixed momentum ($p=0$ GeV/c, $p=10$ GeV/c and $p=50$ GeV/c); (\textbf{bottom}) the spatial diffusion coefficient $D_s2\pi T$ as a function of momentum at fixed temperature ($T=154$ MeV, $T=350$ MeV, $T=550$ MeV). The grey area refers to the prior range before calibration, while the red region refers to the posterior range after calibrating on experimental data. The black dashed line refers to the diffusion coefficient from a leading order pQCD calculation, the red lines are the parametrized diffusion coefficient using the median value of the posterior parameter distributions.}
\end{figure*}

The main results of a Bayesian analysis are the posterior probability distributions of the input parameters $(\alpha, \beta, \gamma)$. The posterior distributions are given in Fig.~\ref{fig:corners_individual} and ~\ref{fig:corners_all}, where the results of 5 independent Bayesian analyses are presented as histograms. Each analysis follows the same procedure described in the previous few sections but calibrates on  different sets of experimental data, and is shown using different colors. For examples, in Fig.~\ref{fig:corners_individual} the blue histograms represent the distributions calibrated on the 5.02 TeV PbPb collision data, the green ones represent the calibration on the 2.76 TeV PbPb data, and the yellows ones represent the calibration on the 200 GeV AuAu data. In Fig.~\ref{fig:corners_all} the magenta histogram corresponds to the analysis using the data from two collision energies at the LHC simultaneously, while the red ones denote the analysis using all the observable across three different systems and listed in Tab.~\ref{tab:obs}. In each figure, the histograms along the diagonal are the marginal posterior probability distributions of each parameter ($\alpha, \beta, \gamma$), with all the other parameters integrated out. The off-diagonal contour plots are the joint distributions which show the correlations among pairs of the parameters. 

Figure~\ref{fig:corners_individual} and~\ref{fig:corners_all} indicate that parameter $\alpha$ is well constrained, peaking around $(1.5\sim 2.0)$ for all analyses. This parameter determines the diffusion coefficient $D_s2\pi T$ at 0 momentum near $T_{\mathrm{c}}$. The slope parameter $\beta$ is poorly contrained, although a negative correlation between $\alpha$ and $\beta$ is observed. Parameter $\gamma$ controls the ratio between the linear component and the pQCD component. As shown in Fig.~\ref{fig:corners_individual}, the distribution for $\gamma$ extracted from AuAu collisions at 200 GeV favors a slightly smaller value than that from the LHC energies, indicating a stronger contribution from the linear component and a slower convergent to the pQCD results in AuAu collision. For the combined analysis of the LHC energies and all three energies, $\gamma$ peaks around $(0.25 \sim 0.3)$. We conclude that for momenta range between $(10-20)$ GeV/c, the linear component and pQCD component of the diffusion coefficient are comparable to each other and that the pQCD contribution to the diffusion coefficient will only dominate at momenta above 20 GeV. The momentum range is approximated using $1/\gamma^2$, as $\gamma^2 p=1$ is the momentum region where linear and pQCD component contributes equally.

The width of the posterior distributions is affected by the uncertainty we have applied in the analysis. The smaller the uncertainty, the stronger the constraint, and therefore a narrower width of the posterior distributions. This may explain why among the three different collision systems, the posterior distributions from 5.02 TeV PbPb collisions generally show a smaller width and the combined calibration is mostly driven by the higher precision data from PbPb 5.02 TeV collisions. Higher precision experimental data and a better understanding of the theoretical uncertainties will yield a significantly better constraint on the parameters.

\subsection{\label{sec:D2piT} Heavy Quark Diffusion Coefficient}

\begin{figure*}
\includegraphics[width=0.9\textwidth]{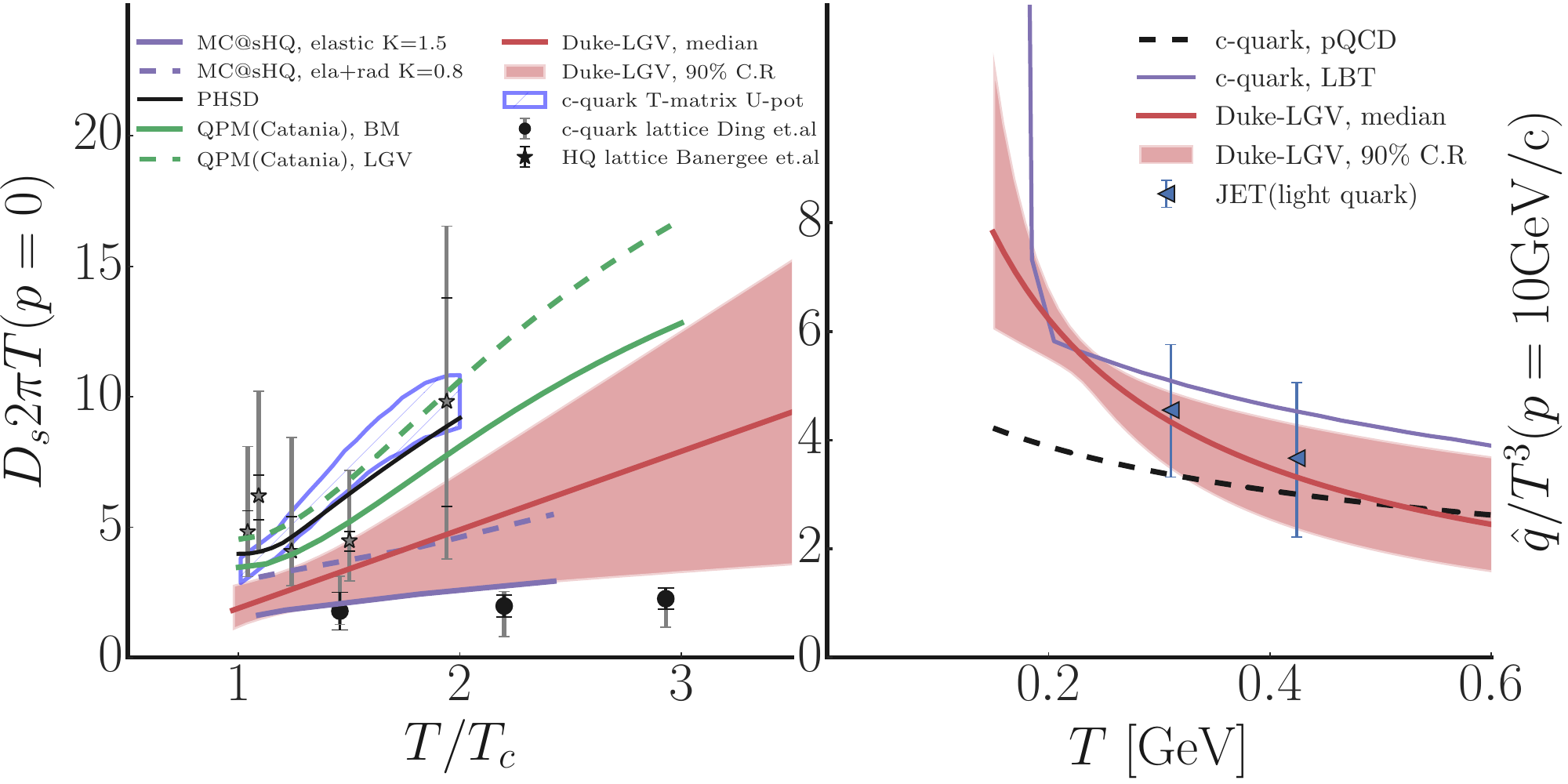}   
\caption{\label{fig:D2piT_compare}(Color online) Comparison of the heavy quark diffusion coefficients across multiple approaches available in the literature. (\textbf{left}) spatial diffusion coefficient at zero momentum $D_s2\pi T(p=0)$. (\textbf{right}) momentum diffusion coefficient $\hat{q}/T^3$ at $p=10$ GeV.}
\end{figure*}

Having established the posterior distribution of the parameters $\alpha, \beta$ and $\gamma$ we can now utilize the parametrization of the spatial diffusion coefficient Eqn.~(\ref{eqn:D2piT}), to extract the posterior range of this quantity. Figure~\ref{fig:D2piT} displays the estimate of the spatial diffusion coefficient, as a function of temperature and momentum respectively. The gray area represents the prior range before the calibration, while the red area is the posterior estimate extracted from the combined analysis with 90\% credibility. Here we use the result of the combined analysis that calibrates on all the three systems. We note, however, the posterior ranges of $D_s2\pi T$ do not differ much between different analysis, even though the posterior distributions of the parameters $(\alpha,\beta,\gamma)$ show some deviation. The dashed black lines represent the diffusion coefficient calculated in leading order pQCD. The solid red lines depict the diffusion coefficient using the median value of the parameters from the posterior distributions. 

On the upper panel of Fig.~\ref{fig:D2piT}, the diffusion coefficient is plotted as function of temperature for 3 different values of the heavy quark momentum ($p=0$ GeV/c, $p=15$ GeV/c and $p=50$ GeV/c): for $p=0$ GeV/c, the diffusion coefficient is solely determined by the linear component, with the $D_s2\pi T (p=0) \sim 1-3$ around $T_{\mathrm{c}}$, (which is the range of parameter $\alpha$ for 5-95\% percentiles). The temperature dependence of $D_s2\pi T$ is not far remote from a simple linear relationship with positive slope. In addition, we notice that the 90\% credibility range suffers the least uncertainties in a temperature range around $T \sim 200 - 250$ MeV, which we argue, is approximately the average temperature that heavy quarks experience during their propagation path. At higher temperature, the posterior range of the spatial diffusion coefficient broadens. A likely reason for this trend is due to the short amount of time the bulk matter retains this high temperature at the beginning of the system's evolution. As the system expands quickly, it rapidly cools, leaving only a short period of time for the heavy quarks to interact with the medium at that temperature, and therefore less information can be obtained at high temperatures.

On the lower panel of Fig.~\ref{fig:D2piT} we explore the momentum dependence of the diffusion coefficient for three different temperatures ($T=150$ MeV, $T=350$ MeV and $T=550$ MeV). As the heavy quark momentum increases, the uncertainties of the posterior range decrease. At high momentum, the diffusion coefficient reflects that of the pQCD calculation, which is obtained from $2\rightarrow 2$ process. The only freedom left in the parameterization is the prefactor $\frac{(\gamma^2 p)^2}{1 + (\gamma^2 p)^2}$,which varies only little for high momenta. In the low momentum region, the parameterized diffusion coefficient shows completely different behavior from the pQCD calculation, which can be only the result of the non-negligible contribution from non-perturbative effects, which are clearly needed in order to obtained a realistic description of the heavy quarks at low and intermediate momentum.

\begin{figure*}
\includegraphics[width=1.0\textwidth]{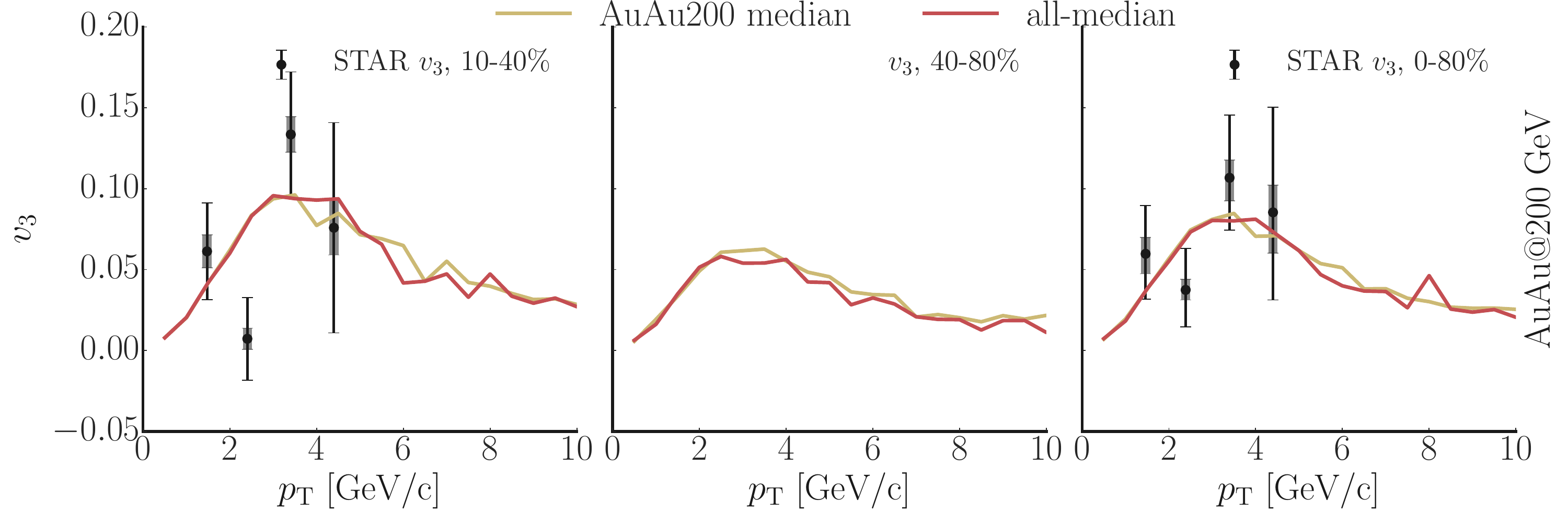}  
\includegraphics[width=1.0\textwidth]{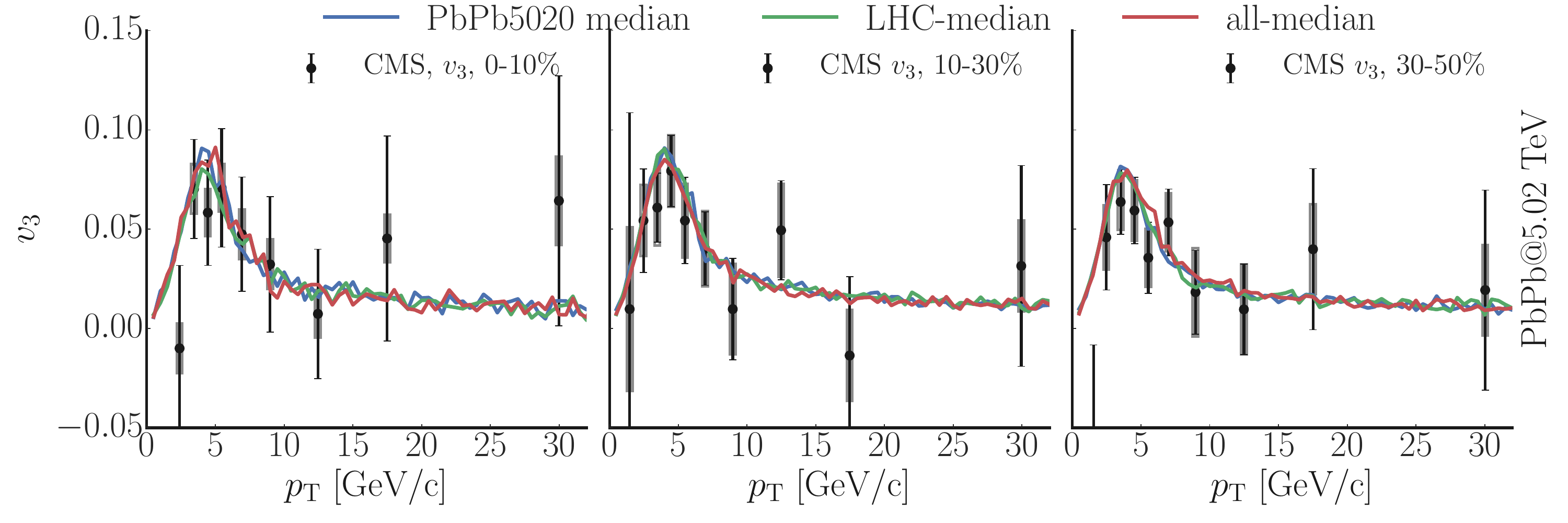}  
\caption{\label{fig:median_v3}(Color online)$D$-meson $v_3$ predicted by the improved Langevin model at AuAu collisions at 200 GeV and PbPb collisions at 5.02 TeV, compared to experimental data measured by STAR~\cite{STAR2017:Dmeson} and CMS~\cite{CMS:2016jtu}. The input parameters ($\alpha, \beta,\gamma$) were set to the median value of the posterior distribution for the AuAu at 200 GeV dataset (AuAu200 median), PbPb at 5.02 TeV dataset (PbPb5020 median), the combined LHC energy dataset (LHC-median), and the combined data set for all three systems (all-median).}
\end{figure*}

In Fig.~\ref{fig:D2piT_compare} we compare our estimate of the diffusion coefficient with the coefficient used or calculated by a number of other models in the market~\cite{Song:2015ykw,Scardina:2017ipo,Riek:2010fk,Gossiaux:2008jv,Cao:2016gvr,Burke:2013yra} as well as with lattice QCD calculations~\cite{Ding:2012sp,Banerjee:2011ra}. The left frame shows the temperature dependence of the spatial diffusion coefficient at 0 momentum $D_s2\pi T(p=0)$. Our analysis is consistent with lattice QCD calculations within the uncertainties currently inherent in lattice QCD calculations. Although the diffusion coefficients used in different models are rather different, they all have a minimum near $T_\mathrm{c}$ with a value range of $D_s2\pi T(p=0)=1-7$.

On the right frame, we compare our charm quark transport coefficient $\hat{q}$ at $p=10$ GeV/c with the results from the LBT model and the JET collaboration~\cite{Cao:2016gvr,Burke:2013yra}. The transport coefficient $\hat{q}$ is roughly comparable with those two. A detailed investigation on the causes of the differences in predicted or calculated transport coefficients by the different approaches is of great interest, but beyond the scope of this work.  We should note that in order to make a valid apple-to-apple comparison among different models, especially for those transport approaches that depend on the surrounding matter, it would be essential for the heavy quarks to propagate in the same QGP medium evolution, which is not the case for this particular comparison. For example, the PHSD group~\cite{Song:2015ykw} explores a non-equilibrium microscopic description of the bulk evolution, the Catania group~\cite{Scardina:2017ipo} describes the medium by solving a relativistic Boltzmann equation for light partons, while the others use a hydrodynamic description. Additionally the properties of the medium and the choice for the equation of state will affect the extraction of the diffusion coefficients as well. 



\subsection{\label{sec:v3}Model validation: triangular flow}

The robustness and quality of our description of in-medium heavy-quark dynamics and our extraction of the heavy quark diffusion coefficient can be tested by making predictions of observables that have not been part of the calibration. Here, we focus on  measurements of higher order flow cumulants for D mesons, in particular $v_3$, which has been predicted as further valuable HQ observables in~\cite{Nahrgang:2014vza}. First measurements of the D meson $v_3$ have recently become available by the CMS and STAR collaborations~\cite{CMS:2016jtu,STAR2017:Dmeson}.
We calculate the $D$-meson $v_3$ in PbPb collisions at 5.02 TeV using the median values of $\alpha, \beta$ and $\gamma$ as determined by our calibration and compare our results to experimental measurements in Fig.~\ref{fig:median_v3}. Each frame contains three results from the improved Langevin model calculation stemming from our different calibrations: utilizing the 200 GeV AuAu data, utilizing the 5.02 TeV PbPb data, utilizing both LHC data sets, and utilizing the combined datasets from all three collision energies and systems. As we have mentioned before, the posterior diffusion coefficients using the median values do not differ much from each other in the different analyses. The robustness of the analysis has been confirmed in Fig.~\ref{fig:median_v3}, where we show that the $D$-meson $v_3$ in AuAu collisions and in all three measured centrality bins in PbPb collisions as predicted by our calibration agrees very well with the data, irrespective of the dataset used to determine the diffusion coefficient.  

\section{Conclusion}
In summary, we have applied state-of-the-art Bayesian methodology to systematically extract the heavy quark diffusion coefficient from a model-to-data analysis of our improved Langevin model for the in-medium heavy quark evolution. By calibrating to the experimental data of $D$-meson $R_{\mathrm{AA}}$ and $v_2$ measured in AuAu collisions at 200 GeV, PbPb collisions at 2.76 TeV and 5.02 TeV, we are able to extract a posterior range of the diffusion coefficient. Our analysis is compatible with lattice QCD calculations within uncertainties that are inherent in the lattice calculations. With the extracted parameters,  our improved Langevin model has been shown to be able to reproduce the experimental data of $R_{\mathrm{AA}}$ and $v_2$ at both RHIC and the LHC simultaneously, and is able to describe well observables that are not included in the calibration, such as $D$-meson $v_3$.

Our parametrization of the spatial diffusion coefficients combines a linear temperature dependent component -- accounting for a non-perturbative contribution --  and a pQCD component -- calculated from a leading order pQCD approach with a fixed coupling of $\alpha_s=0.3$. It smoothly interpolates between the linear component and the pQCD component, and converges to the pQCD calculation in the large momentum limit.  The spatial diffusion coefficient at zero momentum $D_s2\pi T(p=0)$ varies between 1-3 near $T_{\mathrm{c}}$ and exhibits a positive slope for its temperature dependence above $T_{\mathrm{c}}$. Even at momenta in the range of $10-20$ GeV/c, the non-perturbative contribution can not be ignored.

In future work, we shall improve our treatment of the different sources of uncertainty, both theoretical and experimental, that can affect the outcome of our analysis. In addition, we plan to improve our physics model by taking the pre-equilibrium phase of the reaction explicitly into account \cite{Srivastava:2017bcm} and to apply the model-to-data framework to different dynamical models of heavy quark in-medium evolution, for example a comparison between  Langevin and Boltzmann dynamics. Moreover, this study serves as the first application of a Bayesian model-to-data analysis to the heavy flavor dynamics in heavy-ion collisions and we intend to expand its application to the study of other rare probes as well.

\begin{acknowledgments}
The authors thank Guang-You Qin, Jussi Auvinen, Ralf Rapp, and Weiyao Ke for valuable discussion and positive feedback. This work has used the CPU hours provided by the Open Science Grid, which is supported by the National Science Foundation and the U.S. Department of Energy's Office of Science. Y.X. and S.A.B. have been supported by the U.S Department of Energy under grant DE-FG02-05ER41367 and J.B. has been supported by the National Science Foundation under grant ACI-1550225. S.C is supported by the U.S DoE under grant DE-SC0013460 and the National Science Foundation under grant ACI-1550300.
\end{acknowledgments}


\end{document}